\documentclass[12pt]{article}
\renewcommand{\theequation}{\arabic{section}.\arabic{equation}}

\begin{document}

\begin{flushright}
IPNO/TH 97-01
\end{flushright}
\vspace{1 cm}
\begin{center}
{\large \textbf{Relativistic effects in the pionium lifetime}}
\vspace{1.25 cm}

H. Jallouli\footnote{E-mail: jallouli@ipno.in2p3.fr .}
and H. Sazdjian\footnote{E-mail: sazdjian@ipno.in2p3.fr .}\\
\textit{Division de Physique Th\'eorique\footnote{Unit\'e de Recherche des
Universit\'es Paris 11 et Paris 6 associ\'ee au CNRS.},
Institut de Physique Nucl\'eaire,\\
Universit\'e Paris XI,\\
F-91406 Orsay Cedex, France}\\
\end{center}
\vspace{1.25 cm}

\begin{center}
{\large Abstract}
\end{center}

The pionium decay width is evaluated in the framework of chiral perturbation
theory and the relativistic bound state formalism of constraint theory.
Corrections of order $O(\alpha)$ are calculated with respect to the
conventional lowest-order formula, in which the strong interaction amplitude 
has been calculated to two-loop order with charged pion masses. Strong 
interaction corrections from second-order perturbation theory of the bound 
state wave equation are found to be of the order of 0.4\%. Electromagnetic 
radiative corrections, due to pion-photon interactions, are estimated to be 
of the order of $-0.1\%$. Electromagnetic mass shift insertions in internal 
propagators produce a correction of the order of $0.3\%$. The correction due 
to the passage from the strong interaction scattering amplitude evaluated 
with the mass parameter fixed at the charged pion mass to the amplitude
evaluated with the mass parameter fixed at the neutral pion mass is found 
to be of the order of 6.4\%.
\par

\newpage

\section{Introduction}
\setcounter{equation}{0}

The $\pi\pi$ scttering amplitude \cite{w1} represents one of the main
quantities that allow confrontation of predictions of chiral 
perturbation theory \cite{w2,gl1,gl2} with experiment. Unfortunately,
the lack of direct low energy data forces one to reconstruct the low
energy scattering amplitude from extrapolations \cite{p,fp,sch} from
high energy data \cite{em,hh,o} and from indirect information coming 
from $K_{l4}$ decay \cite{r}, at the price of increasing error bars on 
numerical values. The 20\% uncertainty of the ``experimental'' value
of the isospin zero $S$-wave scattering length, $a_0^0=0.26\pm 0.05$
\cite{fp,nd,g,mp}, does not allow one to draw a clear-cut conclusion
when the latter is compared with the theoretical prediction of standard 
chiral perturbation theory, which is 0.20 to the one-loop order \cite{gl1}
and 0.217 to the two-loop order \cite{bcegs}.
\par
From this viewpoint, the DIRAC experiment, which will be realized 
at CERN in the near future and which aims at measuring the pion scattering 
lengths from the lifetime of pionium ($\pi^+\pi^-$ atom) decaying
into $\pi^0\pi^0$ \cite{dir,nm}, might provide a decisive improvement for
the above comparison. (Results on previous experiments are presented 
in Ref. \cite{af}.)\par
On the other hand, it was noticed \cite{fss} on theoretical grounds
that the standard formulation of chiral perturbation theory does not
provide the most general basis for an interpretation of the 
experimental results within the framework of QCD. It is generally assumed 
that the quark condensate in the chiral limit, $-<0|\overline qq|0>_0
/F_{\pi}^2$, has a mass scale of the order of 1 GeV, typical of massive
hadron masses \cite{gl1,gmorgw}. This hypothesis, while apparently
natural, has not yet received a direct experimental confirmation.
In this connection, a generalized form of chiral perturbation theory
was proposed \cite{fss,ssf}, in which the above hypothesis is relaxed and 
the quark condensate itself appears as an additional expansion parameter, 
allowing, in certain processes, its experimental evaluation. Thus, the
$\pi\pi$ scattering amplitude $A(s|t,u)$ becomes at lowest order $O(p^2)$:
\begin{eqnarray} \label{1e1}
A(s|t,u)&=&\frac{1}{F_0^2}(s-2\hat mB_0),\nonumber \\
B_0&=&-<0|\overline qq|0>_0/F_0^2, \ \ \ \ \hat m=\frac{m_u+m_d}{2}.
\end{eqnarray}
$F_0$ is the pion decay constant $F_{\pi}$ in the chiral limit 
($F_{\pi}= 92.4$ MeV), $m_u$ and $m_d$ are the 
masses of the quarks $u$ and $d$; Zweig rule violating effects have been
neglected. The quark condensate parameter $2\hat m B_0$ is expected to 
take values between 0. and $m_{\pi}^2$, the latter value reproducing the
standard predictions of chiral perturbation theory. 
\par
For $2\hat m B_0 \simeq \frac{1}{2}m_{\pi}^2$ for instance, 
one finds for the scattering 
length $a_0^0$, including one-loop effects \cite{kms,kmsf}, the value
0.27, which lies 35\% away from the standard value 0.20 and closer to 
the experimental value 0.26. This example underlines more acutely the
necessity of disposing of precise experimental informations on the pion
scattering lengths to be able to distinguish between various chiral
symmetry breaking schemes.
\par
Coming back to the pionium lifetime, its lowest-order expression was 
established long ago in the nonrelativistic limit by various methods 
\cite{dgbth,up,t,bnnt}:   
\begin{equation} \label{1e2}
\frac{1}{\tau_0} = \Gamma_0 = \frac{16\pi}{9}\sqrt
{\frac{2\Delta m_{\pi}}
{m_{\pi^+}}} \frac{(a_0^0-a_0^2)^2}{m_{\pi^+}^2} |\psi_{+-}(0)|^2,\ \ \
\ \ \Delta m_{\pi}=m_{\pi^+}-m_{\pi^0},
\end{equation}
where $a_0^I$ is the (dimensionless) $S$-wave scattering length in the 
isospin $I$ channel, usually evaluated in the literature with the charged
pion mass, and $\psi_{+-}(0)$ is the wave function of the pionium 
at the origin (in $x$-space). Characteristics of the pionium have been
discussed in Ref. \cite{eil} and the relevance of its lifetime for
determining chiral symmetry breaking parameters has been outlined in
Ref. \cite{bpt}.
While the above formula provides a relationship
between the pionium lifetime and the pion scattering lengths, it is 
desirable, for a precise theoretical interpretation of the experimental 
result, to have a knowledge of the possible corrections to it. This
question was addressed recently in Refs. \cite{mrw,skg,lr,k}.
\par
The purpose of the present paper is to evaluate the corrections to 
formula (\ref{1e2}) in the framework of $SU(2)\times SU(2)$ chiral
perturbation theory. Apart from relativistic kinematic and mass shift 
corrections, they can be grouped into four categories: 
i) Corrections coming from second-order perturbation theory in the
bound state wave equation. ii) Contributions originating from the 
electromagnetic radiative corrections due to pion-photon interactions. 
iii) Contributions coming from the electromagnetic mass shift corrections,
due to quark-photon interactions and acting through  insertions of the
$O(e^2p^0)$ mass shift lagrangian term in pion internal propagators.
iv) Mass shift corrections with respect to the strong interaction 
amplitude evaluated with the charged pion mass.
\par
The evaluation of the pionium bound state energy shift is done in the
framework of constraint theory relativistic wave equations \cite{ct},
which can be considered as a variant of the quasipotential approach
\cite{qp,lp} and has been shown to provide a means of a covariant treatment 
of the QED bound state problem \cite{js}.
The above corrections to the bound state 
energy shift are evaluated to the relative leading order $O(\alpha)$, 
where $\alpha$ is the QED fine structure constant, the calculations being 
generally done to one loop (globally with respect to the strong and
electromagnetic interactions).
\par
Our results are the following. 
The corrections of the first type are found to be of the order of
0.4\%. The corrections of the second type are shown to be free of infra-red
enhancement and are estimated to be of the order of $-0.1\%$.
The corrections of the third type are estimated to be of the order of 
$0.3\%$. The corrections of the fourth type are found to be of the order of 
6.4\%.
\par
The plan of the paper is the following. In Sec. 2, the properties of the 
constraint theory bound state equation are briefly sketched.
In Sec. 3, the latter formalism is adapted to the case of the
coupled channels of the $\pi^+\pi^-$ and $\pi^0\pi^0$ systems. 
In Sec. 4, the pionium lifetime expression in first-order of perturbation 
theory with respect to the strong interactions with pion mass shift is 
established.
In Sec. 5, the corrections due to second-order perturbation theory
in the bound state equation are evaluated. In Sec. 6, the radiative 
corrections due to the pion-photon interaction are evaluated to one-loop 
order in the tree appriximation of the strong interactions. 
In Sec. 7, the electromagnetic mass shift corrections are calculated. 
A summary of the results is presented in Sec. 8. Several details of 
calculations are presented in Appendices A-E.

\par

%\newpage

\section{The constraint theory bound state equation}
\setcounter{equation}{0}

The Bethe--Salpeter equation \cite{sbgmln}, which is the basic 
bound state equation in quantum field theory, has been revealed 
inadequate for quantitative calculations with \textit{covariant}
propagators. Two typical drawbacks are the following. In the 
nonrelativistic limit, the one-photon (or one-particle) exchange
diagram yields relativistic corrections of order $1/c$, instead of
$1/c^2$ \cite{bcm}. In spectroscopic calculations, two-photon exchange
diagrams yield spurious infra-red logarithmic singularities \cite{l}.
These effects are cancelled only with the inclusion of higher order
diagrams, a feature that enormously complicates the use of the equation
in perturbation theory.
\par
In practice, the Bethe--Salpeter equation has been used in QED in
the Coulomb gauge, which is a noncovariant gauge. Because of the
instantaneous nature of the dominant part of the photon propagator,
one is able to transform the original four-dimensional equation into
a three-dimensional one and to avoid the previous difficulties 
\cite{bybrbr}. However, the latter gauge has its own limitations. It
neccesitates a different treatment of exchanged photons and of 
photons entering in radiative corrections. Furthermore, additional
complications arise when QED is mixed with other interactions, where 
already covariant propagators are present.
\par
In this respect, the wave equations obtained in the framework of 
constraint theory \cite{ct} have been shown to provide a satisfactory 
answer to the requirement of a covariant treatment of perturbation theory
in the bound state problem \cite{js}.
\par
In constraint theory the relative energy variable is eliminated by
means of a constraint equation. For a two-particle system this is 
generally chosen in the form:
\begin{equation} \label{2e1}
C(P,p) \equiv  (p_1^2-p_2^2) - (m_1^2-m_2^2) \approx  0 .
\end{equation}
\par
The two-particle Green's function is projected on this hypersurface 
and then iterated around it. At its pole positions, the latter not being
affected by the projection operation (\ref{2e1}), one establishes a 
three-dimensional eigenvalue equation that takes the form:
\begin{equation} \label{2e2}
\widetilde g_0^{-1} \Psi = -\widetilde V \Psi ,
\end{equation}
where $\widetilde g_0^{-1}$ is the wave equation operator, which will be
specified below, and $\widetilde V$ is the potential, related to the
renormalized off-mass shell scattering amplitude $T$ by a 
Lippmann--Schwinger type equation:
\begin{equation} \label{2e3}
\widetilde V = \widetilde T + \widetilde V\widetilde g_0\widetilde T ,
\ \ \ \ \ \ \ \widetilde T = \frac{i}{2\sqrt{s}} T\big \vert_C ,
\end{equation}
where the index $C$ denotes the use of constraint (\ref{2e1}) (on the
external lines of $T$) and $s=(p_1+p_2)^2$. ($T$ is defined as the 
amputated four-point connected Green's function multiplied by the wave
function renormalization factors of the external particles.) The amplitude 
$\widetilde T$ contains the usual
Feynman diagrams, where the external particles are submitted to the
constraint $C$. The second term in the right-hand side of the first of Eqs. 
(\ref{2e3}) generates an iteration series, the diagrams of which are
called ``constraint diagrams'', where the integrations, because of the
presence of the factor $\widetilde g_0$, are three-dimensional, taking
into account constraint $C$. 
\par
\textit{As long as perturbation theory is 
concerned, Eq. (\ref{2e2}) is equivalent in content to the exact
Bethe--Salpeter equation, with, however, a different arrangement of 
the perturbation series.} The constraint theory wave
function $\Psi$ is related to the Bethe--Salpeter wave function $\Phi$
by means of the projection of the latter on the constraint hypersurface 
(\ref{2e1}), but the explicit form of this relationship will not be
needed in the present work. Once Eq. (\ref{2e2}) is solved with the exact
potential $\widetilde V$, the Bethe--Salpeter wave function $\Phi$ can be
reconstructed, through the iteration procedure, in terms of the constraint
theory wave function $\Psi$ and can be shown to satisfy the 
Bethe--Salpeter equation with the exact irreducible kernel $K$ and the 
same energy eigenvalue as that of Eq. (\ref{2e2}) \cite{js}.
\par
The choice of the operator $\widetilde g_0$ is not unique in principle, 
but in practice it is made on the basis of several natural 
requirements. The choice that is made below satisfies the following
four properties: i) Correct nonrelativistic limit (Schr\"odinger 
equation). ii) Correct one-body limit. When one of the masses becomes 
infinite, one recovers the Dirac or Klein-Gordon equation in the presence
of the static potential. iii) Correct hermiticity and unitarity
properties. Potential $\widetilde V$ is an irreducible kernel, in the
sense that it is free of singularities in the $s$-channel, at least in the
elastic unitarity region: the constraint diagrams cancel the singularities
of the reducible diagrams of $\widetilde T$. iv) Correct QED spectroscopy.
In particular, the constraint diagram contributions remove all spurious
singularities (in the bound state region) coming from $\widetilde T$. More
generally, the leading effect of the sum of all $n$-photon exchange
diagrams (in the absence of radiative corrections) is of order 
$O(\alpha^{2n})$, where $\alpha$ is the fine structure constant.
\par
When constraint (\ref{2e1}) is used, the Klein-Gordon operators of
particles 1 and 2 become equal:
\begin{equation} \label{2e4}
H_0 \equiv  (p_1^2-m_1^2)\big \vert _C  = (p_2^2-m_2^2)\big
\vert _C  = \frac {P^2}{4} - \frac {1}{2} (m_1^2+m_2^2) +
\frac {(m_1^2-m_2^2)^2}{4P^2} + p^{T2} .
\end{equation}
We use the notations:
\begin{equation} \label{2e5}
P = p_1 + p_2 ,\ \ \ p = \frac {1}{2}(p_1-p_2) ,\ \ \ 
X = \frac {1}{2}(x_1+x_2) ,\ \ \ x = x_1-x_2 ,
\end{equation}
and the decompositions of four-vectors into transverse and longitudinal
vectors with respect to $P$:
\begin{eqnarray} \label{2e6}
q_{\mu}^T &=& q_{\mu} - \frac {q.P}{P^2} P_{\mu} ,\ \ \ 
q_{\mu}^L = (q.\hat P) \hat P_{\mu} ,\ \ \ \hat P_{\mu}
= \frac {P_{\mu}}{\sqrt {P^2}} ,\ \ \ q_L = q.\hat P\ ,
\nonumber \\
P_L &=& \sqrt {P^2} ,\ \ \ \ r = \sqrt {-x^{T2}} .
\end{eqnarray}
\par
For two spinless particles, $\widetilde g_0$ is chosen to be:
\begin{equation} \label{2e7}
\widetilde g_0 = -\frac{1}{H_0+i\epsilon} ,
\end{equation}
up to possible finite renormalizations related to the finite parts of the 
individual particle propagator renormalizations; they will not show up,
however, to the approximations used throughout this work.
Furthermore, $\widetilde g_0$ undergoes a finite multiplicative
renormalization by a constant factor $(1+\gamma_1)$ due to the off-mass 
shell treatment of the Lippmann--Schwinger type equation (\ref{2e3}). The
constant $\gamma_1$ appears from the requirement that the only 
$O(1/r)$ terms in the QED potential come from the one-photon
exchange diagram. Its presence amounts to multiplying the potential
$\widetilde V$ by $(1+\gamma_1)$ and continuing the use of expression 
(\ref{2e7}) for $\widetilde g_0$, the constant $\gamma_1$ allowing the
cancellation of a spurious $O(\alpha^3)$ term \cite{js}. The use of this 
finite multiplicative constant, which tends to improve the perturbative 
expansion of the potential, should not, however, have an influence on 
physical quantities (in analogy with the presence of wave function 
renormalization constants).  
\par
Since in the present work we are interested by corrections of order
$O(\alpha)$ to the pionium bound state energy, we can from the start
consider the pure QED potential in its nonrelativistic limit (Coulomb
potential) and use the corresponding nonrelativistic wave functions
for the zeroth-order approximations. The pure QED corrections in the
channel $\pi^+\pi^--\pi^+\pi^-$, being of
order $O(\alpha^2)$ \cite{js,n}, will not be considered further.
\par 
The nonrelativistic Coulomb potential is here:
\begin{equation} \label{2e8}
V_{Coul.} = -2\mu\frac{\alpha}{r} , \ \ \ \ \ \ \mu = 
\frac{m_1m_2}{m_1+m_2}.
\end{equation}
\par
The rest of the potential in Eq. (\ref{2e3}) will be treated as a 
perturbation. It contains the strong interaction part of the $\pi\pi$
interaction, as well as the interference part between strong and
electromagnetic interactions.
\par
In general, potential $\widetilde V$ being energy dependent, the scalar
product of wave functions has a more complicated kernel than in the
energy independent case \cite{js}. The perturbation theory formulation
in the case of energy dependent potentials can be found in Ref. \cite{lp}
(valid for four- and three-dimensional equations). However, since in the
present work the zeroth-order potential is the energy independent Coulomb
potential (\ref{2e8}), the scalar product that should be used in the
perturbative calculations is the usual nonrelativistic one. Energy factors,
present in higher-order potentials, should then be expanded around their
zeroth-order values.
\par
In the rest of this work we shall use, for the evaluation of the importance
of various terms, the infra-red counting rules of the QED bound state
system. Let, for a given process $1+2\rightarrow 3+4$, $s,\ t,\ u$ be
the Mandelstam variables: $s=(p_1+p_2)^2$, $t=(p_1-p_3)^2$, $u=
(p_1-p_4)^2$. We also define the (c.m.) momentum operators:
\begin{eqnarray} \label{2e9}
b_{ab}^2(s) &=& \frac{s}{4}-\frac{1}{2}(p_a^2+p_b^2)
+\frac{(p_a^2-p_b^2)^2}{4s}= -p^{T2}
\ \ \ \ (a,b=1,2\  \mathrm{or}\ 3,4),\nonumber \\
b_{0,ab}^2(s) &=& \frac{s}{4}-\frac{1}{2}(m_{\pi^a}^2+m_{\pi^b}^2)+
\frac{(m_{\pi^a}^2-m_{\pi^b}^2)^2}{4s}
=p_{aL}^2-m_{\pi^a}^2=p_{bL}^2-m_{\pi^b}^2. 
\end{eqnarray}
In the pionium state ($\pi^+\pi^-$), the deviation of $s$ from the threshold 
value $4m_{\pi^+}^2$ is of order $O(\alpha^2)$. The quantities
$b_{+-}^2(s)$, $b_{00}^2(s)$, $b_{0,+-}^2(s)$, $t$ and $u$ are of order 
$O(\alpha^2)$. The quantity $b_{0,00}^2$ of the $\pi^0\pi^0$ system at the 
same energy is of order $O(\Delta m_{\pi}/m_{\pi})$.
\par

%\newpage

\section{Wave equations of the \mbox{\boldmath $\pi^+\pi^-$} and
\mbox{\boldmath $\pi^0\pi^0$} systems}
\setcounter{equation}{0}

In order to deal with the specific sectors of the $\pi^+\pi^-$ and
$\pi^0\pi^0$ systems, we have to enlarge the spaces of potentials and
wave functions considered in Sec. 2.  
We introduce a two-component wave function $\Psi$ as:
\begin{equation} \label{4e3}
\Psi = \left( \begin{array}{c}
\Psi_{+-}\\ \Psi_{00} \end{array} \right)
\end{equation}
and define the potential $\widetilde V$ in matrix form in the corresponding
space:
\begin{equation} \label{4e4}
\widetilde V = \left( \begin{array}{rc}
V_{+-,+-} & V_{+-,00}\\
V_{00,+-} & V_{00,00}  \end{array}  \right) .
\end{equation}
\par
The constraint propagator $\widetilde g_0$ [Eqs. (\ref{2e2}),
(\ref{2e3}) and (\ref{2e7})] is now:
\begin{equation} \label{4e5}
\big[\widetilde g_0\big] = \left( \begin{array}{cc}
\widetilde g_{0,+-} & 0\\
0 & \frac{1}{2} \widetilde g_{0,00} \end{array} \right),\ \ \ \ 
\big[\widetilde g_0\big]^{-1} = \left( \begin{array}{cc}
\widetilde g_{0,+-}^{-1} & 0\\
0 & 2\widetilde g_{0,00}^{-1} \end{array} \right) ,
\end{equation}
where the subscripts $+-$ and $00$ have been associated with $\widetilde 
g_0$, in order to take account, when necessary, of the specific
free masses of the systems $\pi^+\pi^-$ and $\pi^0\pi^0$ in formula 
(\ref{2e7}). The factor 1/2 in front of $\widetilde g_{0,00}$ has been 
introduced because of the identity of the particles in the corresponding 
sector.
\par
Let $T_{+-,+-}$, $T_{+-,00}$, $T_{00,+-}$ and $T_{00,00}$ 
(with $\mathcal{M}\equiv -iT$) designate the scattering amplitudes of the 
processes $\pi^+\pi^-\rightarrow\pi^+\pi^-$,
$\pi^0\pi^0\rightarrow\pi^+\pi^-$, etc.. 
We use for our calculations the chiral effective lagrangian 
\cite{gl1,gl2,ssf} in the $SU(2)\times SU(2)$ case. The scattering 
amplitude is obtained from the amputated four-point Green's function
of pseudoscalar densities, multiplied by the corresponding wave function
renormalization factors. We shall use for the field $U$ of the chiral
effective lagrangian the representation
\begin{equation} \label{5ea}
U = \sigma + i\frac{\mbox{\boldmath$\pi$}}{F_0}.\mbox{\boldmath
$\tau$},\ \ \ \ \sigma = \sqrt{1-\frac{\mbox{\boldmath$\pi$}^2}{F_0^2}},
\end{equation}
where \mbox{\boldmath$\tau$} are the Pauli matrices and 
\mbox{\boldmath$\pi$} the pion fields. 
The pseudoscalar densities are defined as $P^a=i\overline q\gamma_5
\tau^aq$ ($a=1,2,3$), where $q$ are the quark fields. Their Green's
functions are obtained by deriving the generating functional with respect 
to the pseudoscalar sources. In the representation (\ref{5ea}) and for
the $O(p^2)$ lagrangian the pseudoscalar densities $P^a$ are proportional
to the fields $\pi^a$ by the common constant factor $2B_0F_0$. In the 
higher-order lagrangian terms more complicated differences arise and 
they should be taken into account. We shall draw Feynman diagrams with 
respect to the pion fields $\pi^a$.
\par
In the strong interaction limit and in the absence of isospin breaking,
the amplitudes $T$ above are related to the conventional amplitudes 
$A(s|t,u)$, etc., with the relations:
\begin{eqnarray} \label{4e1}
-iT_{+-,+-}^{str.} &=& \mathcal{M}_{+-,+-}^{str.} =
A(s|t,u)+A(t|s,u),\nonumber \\
-iT_{+-,00}^{str.} &=& -iT_{00,+-}^{str.} = \mathcal{M}_{+-,00}^{str.} 
= A(s|t,u),\nonumber \\
-iT_{00,00}^{str.} &=& \mathcal{M}_{00,00}^{str.} = 
A(s|t,u)+A(t|s,u)+A(u|t,s),
\end{eqnarray}
and in terms of the isospin invariant amplitudes $T^{(I)}$, they are
\cite{p}:
\begin{eqnarray} \label{4e2}
T_{+-,+-}^{str.} &=&
\frac{1}{6}T^{(2)}+\frac{1}{2}T^{(1)}+\frac{1}{3}T^{(0)},\nonumber \\
T_{+-,00}^{str.} &=& T_{00,+-}^{str.} = \frac{1}{3}T^{(0)}-
\frac{1}{3}T^{(2)},\nonumber \\
T_{00,00}^{str.} &=& \frac{2}{3}T^{(2)}+\frac{1}{3}T^{(0)}.
\end{eqnarray}
The scattering lengths are defined as:
\begin{equation} \label{4e14}
-iT^{(I)} (s=4m_{\pi}^2) = 32\pi a_{l=0}^I .
\end{equation}
\par
Equations (\ref{2e3}) take now the explicit forms:
\begin{eqnarray} \label{4e6}
V_{+-,+-} &=& \widetilde T_{+-,+-} + V_{+-,+-}\widetilde g_{0,+-}
\widetilde T_{+-,+-} + \frac{1}{2}V_{+-,00}\widetilde g_{0,00}
\widetilde T_{00,+-} ,\nonumber \\
V_{+-,00} &=& \widetilde T_{+-,00} + V_{+-,+-}\widetilde g_{0,+-}
\widetilde T_{+-,00} + \frac{1}{2}V_{+-,00}\widetilde g_{0,00}
\widetilde T_{00,00} ,\nonumber \\
V_{00,+-} &=& \widetilde T_{00,+-} + V_{00,+-}\widetilde g_{0,+-}
\widetilde T_{+-,+-} + \frac{1}{2}V_{00,00}\widetilde g_{0,00}
\widetilde T_{00,+-} ,\nonumber \\
V_{00,00} &=& \widetilde T_{00,00} + V_{00,+-}\widetilde g_{0,+-}
\widetilde T_{+-,00} + \frac{1}{2}V_{00,00}\widetilde g_{0,00}
\widetilde T_{00,00} .
\end{eqnarray}
\par
We isolate from the potential $V_{+-,+-}$ the Coulomb potential part
[Eq. (\ref{2e8})]:
\begin{equation} \label{4e6a}
V_{+-,+-}=V_{Coul.}+\overline V_{+-,+-}.
\end{equation}
The wave equations (\ref{2e2}) then become:
\begin{eqnarray}
\label{4e7}
-\widetilde g_{0,+-}^{-1} \Psi_{+-} &=& \big(V_{Coul.}+
\overline V_{+-,+-}\big) \Psi_{+-} + V_{+-,00}\Psi_{00} ,\\
\label{4e8}
-\widetilde g_{0,00}^{-1} \Psi_{00} &=& \frac{1}{2}V_{00,+-}\Psi_{+-}
+ \frac{1}{2}V_{00,00}\Psi_{00} .
\end{eqnarray}
\par
These two wave equations are characterized by the same eigenvalue $P^2$.
\textit{Whenever not specified, all potentials will be
considered at the pionium ground state c.m. energy.}
At zeroth order (Coulomb potential only), it is:
\begin{equation} \label{4e9}
P_0 = 2m_{\pi^+}-\frac{m_{\pi^+}}{4}\alpha^2 .
\end{equation}
\par
For this value of $P_0$, the operator $\widetilde g_{0,00}$ in Eq.
(\ref{4e8}) has values in the scattering region of the $\pi^0\pi^0$
system and therefore it gives rise to a scattering state for the
wave function $\Psi_{00}$. The energy shift of the pionium is obtained
by first eliminating $\Psi_{00}$ from Eq. (\ref{4e7}) in terms of
$\Psi_{+-}$ through Eq. (\ref{4e8}), using there the boundary condition
that $\Psi_{00}$ is an outgoing wave due entirely to the presence of
$\Psi_{+-}$.
\par
The wave function $\Psi_{00}$ can be expressed in terms of $\Psi_{+-}$ by
iterating Eq. (\ref{4e8}) with its last term (proportional
to $V_{00,00}$). This iteration, where the dominant contribution comes
from the strong interaction sector, when treated globally as a
first-order perturbation, yields in the pionium decay width expression 
the unitarity factor \cite{up,bnnt}
$(1+(2/9)(\Delta m_{\pi}/m_{\pi})(2a_0^2+a_0^0)^2)^{-1}$; the correction 
term to one, being of the order of $10^{-4}$, will be neglected in the 
following. We thus obtain the wave equation:
\begin{equation} \label{4e10}
-\widetilde g_{0,+-}^{-1}\Psi_{+-} = \bigg[V_{Coul.}+\overline V_{+-,+-}
-\frac{1}{2} V_{+-,00,}\widetilde g_{0,00}V_{00,+-}\bigg]\Psi_{+-} .
\end{equation}
\par
To the order of approximations we are working, the potentials 
$\overline V_{+-,+-}$, $V_{+-,00}$ and $V_{00,+-}$ are, in $x$-space,
three-dimensional delta-functions and hence they project all 
multiplicative quantities on their values at the origin. 
The value at the origin of the function $\widetilde g_{0,00}$ is 
calculated by Fourier transformation to momentum space and dimensional 
regularization. Designating
by $\Delta m_{\pi}$ the pion mass difference [Eq. (\ref{1e2})],
one finds (see Appendix A, Eq. (\ref{ae3})):
\begin{equation} \label{4e12}
\widetilde g_{0,00} (r=0) = \frac{i}{4\pi}\sqrt
{\Delta m_{\pi} (m_{\pi^+}+m_{\pi^0})} .
\end{equation}
\par
This term induces an imaginary part to the pionium energy. 
\par

%\newpage

\section{Lowest-order formula with mass shift}
\setcounter{equation}{0}

We shall determine a lowest-order expression for the pionium decay 
width with the inclusion in it of the main part of the pion mass shift. 
This formula will prove useful for the evaluation of the various types of
correction.
It is obtained by treating the last term of Eq. (\ref{4e10})
in first-order of perturbation theory and by keeping in the potential
$V_{00,+-}$ its dominant part, which comes essentially from the strong
interactions. The expression of $V_{00,+-}$ [Eq. (\ref{4e6})] contains
three terms. The second  and third terms are the constraint diagram 
contributions, which will be considered below. The first term is
the scattering amplitude. The latter
can be split into two parts, the strong interaction part and the rest,
which represents the contributions containing electromagnetic effects.
Among the latter, there is one piece which plays a crucial role; it is the
quark-photon (or massive hadron-photon) interaction term in the chiral limit,
which is reponsible of the main part of the pion mass difference 
\cite{dgmly}. It corresponds to the $O(e^2 p^0)$ term of the chiral effective 
lagrangian with the expression $e^2C<QUQU^{\dagger}>$, where $C$ is an 
unknown constant and $Q$ is the quark charge matrix \cite{egpdr,u}.
Using for the field $U$ the representation (\ref{5ea}), one finds:
\begin{equation} \label{5e21}
e^2C<QUQU^{\dagger}> = -\frac{2e^2C}{F_0^2}\pi^+\pi^-,
\end{equation}
which shows that this term induces for the charged pions the mass shift:
\begin{equation} \label{5e22}
\big(\Delta m_{\pi}^2\big)_{q\gamma} = 2e^2\frac{C}{F_0^2} ,
\end{equation}
which is nonvanishing in the chiral limit. Otherwise, it has no effect
on the scattering amplitude in lowest order. (We emphasize that
expression (\ref{1e1}) of the latter is also valid off the mass-shell
and does not depend of any mass-shell prescription.)
Therefore, the quark-photon interaction term (\ref{5e21}) acts essentially 
through insertions in the pion loop propagators, where the charged pion 
masses are replaced by their (almost) physical masses. Because the pionium 
lifetime evaluation is sensitive to the pion mass difference, it is natural 
to incorporate from the start the quark-photon interaction term (\ref{5e21})
and its counterterms of the higher-order lagrangian 
in the strong interaction lagrangian. With this prescription,
the amplitude $\mathcal{M}$ is split into two terms:
\begin{equation} \label{3e16}
\mathcal{M}=\mathcal{M}^{str.+q\gamma}+\mathcal{M}^{em.},
\end{equation}
where $\mathcal{M}^{em.}$ contains all interference terms between
electromagnetic and strong interaction effects, except the lowest-order
quark-photon interaction term and its counterterms, which are included, 
together with the strong interaction terms, in $\mathcal{M}^{str.+q\gamma}$.
\par
We now turn to the evaluation of the two constraint diagrams corresponding
to the last two terms of the expression of the potential $V_{00,+-}$ [Eq.
(\ref{4e6})]. The evaluation of these diagrams is done by first considering 
the potentials and amplitudes $\widetilde T^{str.}$ at the tree level 
(Fig. 1) and calculating the loop integrals with the physical pion masses
(cf. the comments above). The details of the calculations are
presented in Appendix A. It is found that the terms  in the amplitudes
proportional to $t$, $u$, $b^2$ and $b_0^2$ yield, after integration, terms
that are again of the same order and hence can be neglected. The leading
terms are directly obtained by ignoring the above terms in the amplitudes
and integrating over $\widetilde g_0$. One finds:
\begin{eqnarray} \label{3e20}
\widetilde T_{+-,+-}\widetilde g_{0,+-}\widetilde T_{+-,+-} &=&
-\frac{1}{4\pi} \sqrt{-b_{0,+-}^2} (\widetilde T_{+-,+-})^2 ,\nonumber \\
\widetilde T_{+-,00} \widetilde g_{0,00} \widetilde T_{00,+-} &=&
+\frac{i}{4\pi}\sqrt{b_{0,00}^2} \widetilde T_{+-,00}\widetilde T_{00,+-},
\end{eqnarray}
(there are no integrations in the right-hand side,) and so forth for the 
other amplitudes, it being clear that $\widetilde g_{0,+-}$ yields a real
contribution and $\widetilde g_{0,00}$ an imaginary one. By the very 
choice of the constraint propagators $\widetilde g_0$, the imaginary
terms that arise from the constraint diagrams cancel similar terms
coming from loop diagrams contained in the amplitudes $\widetilde T$
(the first terms of the right-hand sides of Eqs. (\ref{4e6})). Therefore,
these amplitudes are replaced by their real parts. Concerning the real part
of the constraint diagram, it is of order $O(\alpha)$ and has the opposite
value of the $O(\alpha)$ part of the deviation of the scattering amplitude
from the $\pi^+\pi^-$ threshold down to the pionium energy; it comes
from the finite non-analytic part of the unitarity loop, the polynomial
parts giving only contributions of order $O(\alpha^2)$ to the above 
deviation. Hence, the real part of the constraint diagram shifts the
real part of the scattering amplitude to its value at the $\pi^+\pi^-$
threshold.
We thus obtain:
\begin{eqnarray} \label{3e21}
V_{+-,+-}^{(0)} &=& \mathcal{R}e\widetilde T_{+-,+-}^{(0)}(s=4m_{\pi^+}^2),
\nonumber \\
V_{00,+-}^{(0)} &=& \mathcal{R}e\widetilde T_{00,+-}^{(0)}(s=4m_{\pi^+}^2),
\nonumber \\
V_{00,00}^{(0)} &=& \mathcal{R}e\widetilde T_{00,00}^{(0)}(s=4m_{\pi^+}^2).
\end{eqnarray}
\par
The above property can also be generalized to the two-loop level of the
strong interaction amplitude. The details of the derivation are 
presented in Appendix B.
\par
The part of the potential $V_{00,+-}$ that will contribute to the
lowest-order expression of the decay width is then provided by the real
part of $\mathcal{M}_{00,+-}^{str.+q\gamma}$ through $\mathcal {R}e 
\widetilde T_{00,+-}^{(0)}$. Defining
\begin{equation} \label{4e13}
P_0 = P_{0R}-i\frac{\Gamma}{2} ,
\end{equation}
one finds the modified lowest-order expression of the pionium decay width:
\begin{equation} \label{3e15}
\frac{1}{\overline \tau_0}=\overline \Gamma_0=\frac{1}{64\pi m_{\pi^+}^2}
\big(\mathcal{R}e\mathcal{M}_{00,+-}^{str.+q\gamma}\big)^2 |\psi_{+-}(0)|^2
\sqrt{\frac
{2\Delta m_{\pi}}{m_{\pi^+}}(1-\frac{\Delta m_{\pi}}{2m_{\pi^+}})},
\end{equation}
where $\mathcal {R}e\mathcal{M}_{00,+-}^{str.+q\gamma}$ is calculated at the 
$\pi^+\pi^-$ threshold. 
($\psi$ is the relative motion part of $\Psi$.)
Furthermore, because the external particle
momenta are subjected to the constraints $p_{1L}=p_{2L}=P_L/2$,
$p_{3L}=p_{4L}=P_L/2$, with $P_L=2m_{\pi^+}$ and the kinetic energy 
operators $|p^{T2}|$ and $|p^{\prime T2}|$ have, in the bound state,
values of the order of magnitude of $O(\alpha^2)$, the mass-shell conditions 
in the above amplitude are:
\begin{equation} \label{5e56}
p_1^2=p_2^2=p_3^2=p_4^2=m_{\pi^+}^2,\ \ \ \ \ s=4m_{\pi^+}^2,\ \ \ t=u=0.
\end{equation}
\par
With respect to the lowest-order formula (\ref{1e2}), formula
(\ref{3e15}) contains two corrections. The first one, which has a kinematic
origin, is included in the square-root term. The second one is included in
$\mathcal{M}^{str.+q\gamma}$, where now, because of the pion mass difference,
a modification occurs from the expression involving the scattering lengths
calculated with the strong interaction amplitude.
\par
In the remaining part of this paper we shall evaluate the 
$O(\alpha)$-corrections to the formula (\ref{3e15}) as well as the 
modifications contained in $\mathcal{M}^{str.+q\gamma}$.
\par

%\newpage

\section{Second-order perturbation theory}
\setcounter{equation}{0}

In this section, we evaluate the effects coming from the second-order
perturbation theory treatment of the strong interaction potential.
At this order, it is the interference of the
last two potential terms of Eq. (\ref{4e10}) that contributes to 
the imaginary part 
of the energy. We have to distinguish here between the contributions of the
discrete and continuum states of the pionium spectrum. We first
consider the discrete spectrum contribution.
\par

\subsection{Contribution of the discrete spectrum}

Designating by $\psi_{+-,n}$ ($n\geq 1$) the zeroth order radial
exitation wave functions of the ground state [$\psi$ is the relative motion
part of $\Psi$] and by $E_n$ the corresponding nonrelativistic energies,
the shift in the decay width is:
\begin{eqnarray} \label{4e15}
\big(\Delta \Gamma\big)_{discr.} &=& -i\frac{2}{m_{\pi}}\widetilde g_{0,00}
(r=0) V_{00,+-}^2\overline V_{+-,+-}\psi_{+-}^2(0)\nonumber \\
& &\ \ \ \ \ \ \ \times \sum_{n=1}^{\infty} \frac{\psi_{+-,n}^2(0)}
{m_{\pi}(E_0-E_n)} .
\end{eqnarray}
\par
Using the nonrelativistic formulas
\begin{equation} \label{4e16}
E_n = -\frac{m_{\pi}\alpha^2}{4(n+1)^2} ,\ \ \ \ 
\psi_{+-,n}^2(0) = \frac{m_{\pi}^3\alpha^3}{8\pi (n+1)^3}\ \ \ \
\ \ (n\geq 0) ,
\end{equation}
one finds:
\begin{equation} \label{4e17}
\big(\Delta \Gamma\big)_{discr.} = -\frac{\alpha}{4\pi}m_{\pi}
\overline V_{+-,+-}\Gamma_0 = \frac{\alpha}{3} (2a_0^0+a_0^2) \Gamma_0 ,
\end{equation}
where in the last expression we have neglected the pion mass difference
and introduced the strong interaction scattering lengths (\ref{4e14}).
\par

\subsection{Contribution of the continuous spectrum}

For the evaluation of the contribution of the continuum states, we introduce
the nonrelativistic wavevector modulus $k=\sqrt{Em_{\pi}}$ and normalize
wave functions as:
\begin{equation} \label{4e18}
\int d^3 \mathbf{x} \psi_{k'}^*(\mathbf{x}) \psi_{k}(\mathbf{x}) =
2\pi \delta(k-k')  .
\end{equation}
\par
The shift in the decay width is then:
\begin{equation} \label{4e19}
\big(\Delta \Gamma\big)_{cont.} = -2\Gamma_0 \overline V_{+-,+-} \int
\frac{dk}{2\pi} \frac{\vert \psi_k(0)\vert^2}{(k^2+m_{\pi}^2\alpha^2/4)} ,
\end{equation}
with \cite{llf}:
\begin{equation} \label{4e20}
\vert \psi_k(0)\vert^2 = \frac{1}{4\pi}\frac{4\pi m_{\pi}\alpha k}
{\big(1-\exp(-\pi m_{\pi}\alpha/k)\big)} .
\end{equation}
\par
It is easily seen that $\big(\Delta \Gamma\big)_{cont.}$ diverges
linearly in the ultra-violet region. In dimensional regularization, 
linear divergences being equivalent to zero, the divergence that survives
in Eq. (\ref{4e19}) is logarithmic:
\begin{equation} \label{4e21}
{\big(\Delta \Gamma\big)_{cont.}}_{\stackrel{{\displaystyle \approx}}
{k\rightarrow\infty}} -2\Gamma_0 \overline V_{+-,+-} \frac{m_{\pi}
\alpha}{4\pi} \ln(\frac{k}{m_{\pi}}) .
\end{equation}
\par
The origin of this divergence is related to the singularity of the 
three-dimensional delta-function that characterizes the potentials in 
$x$-space. In quantum mechanics, potentials that are proportional to 
three-dimensional delta-functions must be regularized through a
renormalization of the coupling constant \cite{bf} or a self-adjoint 
extension of the hamiltonian \cite{j}. In both cases, an additional unknown
parameter appears in the spectrum. On the other hand, a bound state
equation, based on field theory, should not introduce new unknown
parameters into the set of parameters already defined or fixed by the 
field theory itself. This is why one should expect the cancellation of
the divergence (\ref{4e21}) by some other terms present in the potential.
\par
The constraint diagrams corresponding to the process $\pi^+\pi^-\rightarrow
\pi^0\pi^0$ with one photon exchange and two loops (Fig. 2) play this role.
They have overall logarithmic divergences that cancel the one appearing
in $\big(\Delta \Gamma\big)_{cont.}$. Therefore, these two types of
contribution should be considered together.
\par
We now consider the contributions of the above constraint diagrams.
At the vertex $\pi^+\pi^--\pi^0\pi^0$, the amplitude that contributes is
$A(s|t,u)$ [Eqs. (\ref{4e1}) and (\ref{1e1})]; in lowest order, it depends 
only on the variable $s$ and therefore can be factored out of the integrals.
At the vertex $\pi^+\pi^--\pi^+\pi^-$, it is the amplitude 
$A(s|t,u)$$+$$A(t,|s,u)$ that contributes. However, the contribution
of the variable $t$, present in $A(t|s,u)$, is of order $O(\alpha^2)$
when considered in the bound state domain (cf. also Appendix A); therefore, 
it can be omitted
and the amplitude $A(s|t,u)$$+$$A(t|s,u)$ can also be factored out.
We use for the potentials arising from the constraint diagrams
the notation $V_{00,+-}^{C(n,m,p)}$, 
$C$ referring to ``constraint'', $n$ to the number of loops, $m$ to
the number of exchanged photons and $p$ to the number of the constraint
factors $\widetilde g_0$.
\par
We first consider the constraint diagram of Fig. 2-$c$. The corresponding
potential is (in the Feynman gauge):
\begin{eqnarray} \label{4e22}
V_{00,+-}^{C(2,1,2)} &=& e^2 \frac{i}{2\sqrt{s}} V_{00,+-}
\overline V_{+-,+-}
\int \frac{d^3k^T}{(2\pi)^3} \frac{d^3k^{\prime T}}{(2\pi)^3}\nonumber \\
& &\ \ \ \ \times \frac{(2p_1-k^T-k^{\prime T}).(2p_2+k^T+k^{\prime T})}
{\big((p_1-k^{\prime T})^2-m_{\pi}^2+i\epsilon\big)
\big((p_1-k^T)^2-m_{\pi}^2+i\epsilon\big)} \frac{-i}{(k^T
-k^{\prime T})^2+i\epsilon}\nonumber \\
&\equiv& -ie^2V_{00,+-}V_{+-,+-}\int \frac{d^3k^T}
{(2\pi)^3} \frac{1}{(p_1-k^T)^2-m_{\pi}^2+i\epsilon}
\nonumber \\
& &\ \times \bigg\{ (4p_1.p_2+4p^T.k^T)F^{(1)C}(p_1,k^T)
+4(p^T-k^T)^{\mu} F_{\mu}^{(1)C}(p_1,k^T) - R^C(p_1)\bigg\},\nonumber \\
& & \ \ \ 
\end{eqnarray}
with:
\begin{eqnarray} \label{4e23}
F^{(1)C}(p_1,k^T) &=& \frac{i}{2\sqrt{s}}\int \frac{d^3k^{\prime T}}
{(2\pi)^3} \frac{1}{\big((p_1-k^{\prime T})^2-m_{\pi}^2+i\epsilon\big)
\big((k^T-k^{\prime T})^2+i\epsilon\big)}\nonumber \\
&=& F^{(1)C}(p_1-k^T,0)\nonumber \\
&=& \frac{i}{4\sqrt{s}}\frac{1}{4\pi}\frac{2}
{\sqrt{-(p^T-k^T)^2}} \arctan\sqrt{\frac{-(p^T-k^T)^2}{-b_0^2(s)}} ,
\nonumber \\
& & 
\end{eqnarray}
\begin{eqnarray} \label{4e23a}
F_{\mu}^{(1)C}(p_1,k^T) &=& \frac{i}{2\sqrt{s}}\int \frac{d^3k^{\prime T}}
{(2\pi)^3} \frac{k_{\mu}^{\prime T}}{\big((p_1-k^{\prime T})^2-m_{\pi}^2
+i\epsilon\big) \big((k^T-k^{\prime T})^2+i\epsilon\big)}\nonumber \\
&=& k_{\mu}^TF^{(1)C}(p_1-k^T,0)+F_{\mu}^{(1)C}(p_1-k^T,0) ,\nonumber \\
F_{\mu}^{(1)C}(p_1,0) &=& \frac{i}{8s}p_{\mu}^T \sqrt {\frac{s}{b^2}}
\bigg(\big(\frac{b^2-b_0^2}{b^2}\big)\arctan\sqrt{\frac{b^2}{-b_0^2}} 
-\sqrt{\frac{-b_0^2}{b^2}}\bigg) ,
\end{eqnarray}
\begin{equation} \label{4e23b}
\overline R^C(p_1) = \frac{i}{2\sqrt{s}}\int\frac{d^3k^T}{(2\pi)^3}
\frac{1}{(p_1-k^T)^2-m_{\pi}^2+i\epsilon} = -\frac{i}{2\sqrt{s}}
\frac{1}{4\pi}\sqrt{-b_0^2} . 
\end{equation}
[The integrals are finite in dimensional regularization.
We have kept in $F_{\mu}^{(1)C}$ the leading terms only.]
\par
In Eq. (\ref{4e22}), it is only the term proportional to $p_1.p_2$,
through $p_{1L}p_{2L}$, that gives the surviving $O(\alpha)$
contribution. This observation makes also clear that the consideration of
other covariant gauges would also yield the same leading term. Therefore,
the subsequent results are gauge independent.
One obtains:
\begin{equation} \label{4e25}
V_{00,+-}^{C(2,1,2)} = -m_{\pi}\alpha V_{00,+-}\overline V_{+-,+-}\int 
\frac {d^3k^T}{(2\pi)^3} \frac{1}{m_{\pi}^2-p_{1L}^2-k^{T2}} \frac{1}
{\sqrt{-k^{T2}}} \arctan\sqrt{\frac{-k^{T2}}{-b_0^2}} .
\end{equation}
\par
We next consider the two constraint diagrams $a$ and $b$ of Fig. 2.
Since the amplitude $A(s|t,u)$$+$$A(t|s,u)$ has been factored out of 
the integrals, the two diagrams give equal contributions. The
four-dimensional loop corresponds to the electromagnetic radiative 
correction of the scalar vertex. Its expression will be calculated in Sec. 
6 (and Appendix D). It contains, among other terms, a term that is the 
opposite of the
function $F^{(1)C}$ [Eqs. (\ref{4e22})-(\ref{4e23})] we calculated above.
[It is the first term in the expression of $F^{(1)}$ in Eq. (\ref{5e11}),
where $b^2\equiv -p^{T2}$ should now be replaced by $-(p^T-k^T)^2$.]
Its other terms contribute to nonleading effects in $\alpha$.
\par
Therefore, the contribution resulting from the sum of the two
diagrams (a) and (b) of Fig. 2 is:
\begin{equation} \label{4e26}
V_{00,+-}^{C(2,1,1)} = -2V_{00,+-}^{C(2,1,2)} .
\end{equation}
\par
The contribution to the shift in the decay width resulting from the three
constraint diagrams of Fig. 2 is then:
\begin{equation} \label{4e27}
\big(\Delta \Gamma\big)^{C(2,1,1+2)} = 2m_{\pi}\alpha \Gamma_0
\overline V_{+-,+-}\int \frac{d^3\mathbf{k}}{(2\pi)^3} 
\frac{1}{(\mathbf{k^2}+m_{\pi}^2\alpha^2/4)} 
\frac{1}{k} \arctan \big(\frac{2k}{m_{\pi}\alpha}\big),
\end{equation}
where we have replaced $-b_0^2\equiv m_{\pi}^2-p_{1L}^2$ by its
eigenvalue $m_{\pi}^2\alpha^2/4$. 
\par
$\big(\Delta \Gamma\big)^C$ has an ultra-violet divergence that is the
opposite of that of $\big(\Delta \Gamma\big)_{cont.}$ [Eq. (\ref{4e21})].
Therefore, the sum $\big(\Delta \Gamma\big)^C$$+$$\big(\Delta
\Gamma\big)_{cont.}$ is finite. One obtains:
\begin{eqnarray} \label{4e28}
\big(\Delta \Gamma\big)^{C(2,1,1+2)} + \big(\Delta \Gamma\big)_{cont.}
&=& 2m_{\pi}\alpha \Gamma_0 \overline V_{+-,+-} \frac{1}{4\pi}\nonumber \\
& &\ \times\int_0^{\infty} dx\left(\frac{x}{x^2+1}\right)
\left[\frac{2}{\pi}\arctan x + \frac{x}{\pi} - \frac{2}{1-e^{-2\pi/x}}
\right]\nonumber \\
&\simeq& -1.7 m_{\pi} \frac{\alpha}{2\pi} \Gamma_0 \overline V_{+-,+-} 
= 1.1 \alpha (2a_0^0+a_0^2) \Gamma_0 .
\end{eqnarray}
[We have subtracted from $\big(\Delta \Gamma\big)_{cont.}$ its linear
divergence, which is null in dimensional regularization.]
The constraint diagrams have played the role of an effective cut-off
of the divergence of $\big(\Delta \Gamma\big)_{cont.}$. Had we cut the
integral in Eq. (\ref{4e19}) at the value $k=m_{\pi}\alpha$, we would have
found a result close to that of Eq. (\ref{4e28}).
\par
In the previous cancellation mechanism of divergences we used 
dimensional regularization and hence ignored linear divergences. The 
latter are present in conventional calculations of integrals and for
such type of calculation they should be taken into account. The point is 
that in this case the constraint diagrams themselves are linearly divergent 
and these divergences should be isolated and grouped essentially with those 
coming from second-order perturbation theory. It is not difficult to show 
that the linear divergences cancel out among themselves and the physical
results are not affected by their presence. Details of the derivation
are presented in Appendix C. The cancellation mechanism of divergences is 
therefore regularization scheme independent.
\par
The total amount of the strong interaction corrections is given
by the sum of $\big(\Delta \Gamma\big)_{discr.}$ [Eq. (\ref{4e17})] and
$\big(\Delta \Gamma\big)^{C(2,1,1+2)}$$+$$\big(\Delta \Gamma\big)_{cont.}$
[Eq. (\ref{4e28})]. One obtains:
\begin{eqnarray} \label{4e31}
\big(\Delta \Gamma\big)_{str.} &=& \big(\Delta \Gamma\big)_{discr.} +
\big(\big(\Delta \Gamma\big)_{cont.}+\big(\Delta \Gamma\big)^{C(2,1,1+2)}
\big) \nonumber \\
&=& 1.5\alpha (2a_0^0+a_0^2) \Gamma_0.
\end{eqnarray}
\par
The value of $\big(\Delta \Gamma\big)_{str.}$ depends on the values of
the scattering lengths and hence also of the quark condensate parameter
$2\hat m B_0$ [Eq. (\ref{1e1})]. In the standard case ($2\hat mB_0$
$\simeq$$m_{\pi}^2$) \cite{gl1,g} one has 1.5($2a_0^0$$+$$a_0^2$)$\simeq$0.55. 
In the other extreme case ($2\hat mB_0=0$), one has 1.5($2a_0^0$$+$$a_0^2$)
$\simeq$0.9 \cite{kms,kmsf}. Therefore, in all cases the correction
$(\Delta \Gamma/\Gamma_0)_{str.}$ is bounded by the values:
\begin{equation} \label{4e32}
\big(\frac{\Delta \Gamma}{\Gamma_0}\big)_{str.} = \left\{ \begin{array}{ll}
0.55\alpha =0.004 &\ \ \ \ \ \ (2\hat mB_0 = m_{\pi}^2),\\
0.9 \alpha =0.0065 & \ \ \ \ \ \ (2\hat m B_0 = 0). 
\end{array} \right.
\end{equation}
\par

%\newpage

\section{Electromagnetic radiative corrections}
\setcounter{equation}{0}

In this section we calculate the electromagnetic radiative corrections
arising fom pion-photon interaction and contributing to the amplitude
$\mathcal {M}^{em.}$ of the decomposition (\ref{3e16}).
\par
We first evaluate the unrenormalized quantities. 
The lagrangian corresponding to the pion-photon interaction is obtained
by appropriately incorporating in the external vector current of the
chiral effective lagrangian the photon field. Using for the field $U$
the representation (\ref{5ea}), the corresponding lagrangian becomes in 
lowest order:
\begin{eqnarray} \label{5e1}
\mathcal{L} &=& \frac{1}{2}\partial_{\mu}\pi^0\partial^{\mu}\pi^0
+(\partial_{\mu}+ieA_{\mu})\pi^+(\partial^{\mu}-ieA^{\mu})\pi^-\nonumber \\
& & -\frac{1}{2}m_{\pi}^2(\pi^0\pi^0+2\pi^+\pi^-) - \frac{1}{4}
F_{\mu\nu}F^{\mu\nu} - \frac{1}{2(1-\xi)}(\partial.A)^2\nonumber \\
& & +\frac{1}{8F_0^2}\partial_{\mu}(\pi^{02}+2\pi^+\pi^-)\partial^{\mu}
(\pi^{02}+2\pi^+\pi^-)\nonumber \\
& &-\frac{1}{8F_0^2}(2\hat mB_0) \big(\pi^{04}
+4\pi^+\pi^-\pi^0\pi^0 + 4\pi^{+2}\pi^{-2}\big),
\end{eqnarray}
$2\hat m B_0$ being defined in Eq. (\ref{1e1}).
\par
We begin with the self-energy correction, which is (photon tadpole
contributions are null in dimensional regularization):
\begin{eqnarray} \label{5e2}
-i\Sigma(p) &=& -e^2 \int \frac{d^4k}{(2\pi)^4} \frac{\big((2p-k)^2
-\xi [(2p-k).k]^2/k^2\big)}
{\big((p-k)^2-m_{\pi}^2+i\epsilon\big)\big(k^2+i\epsilon\big)}\nonumber \\
&\equiv& -e^2 \bigg[4p^2R^{(1)}(p)+(1-\xi)T(p)-4(1-\xi)p^{\mu}R_{\mu}^{(1)}
(p) -4\xi p^{\mu}p^{\nu}R_{\mu\nu}^{(2)}(p)\bigg],\nonumber \\
& &
\end{eqnarray}
where $\xi$ represents the gauge parameter. The definitions and expressions 
of the functions $R$ and $T$ are presented in Appendix D. 
The expression of the unrenormalized self-energy is given in Eq. 
(\ref{be1}). The unrenormalized Green's function of the charged 
pseudoscalar density becomes:
\begin{equation} \label{5e7}
G^+(p) = \frac{i(2B_0F_0)^2}{p^2-(m_{\pi}^2+(\Delta m_{\pi}^2)_{\pi\gamma}) 
+i\epsilon} \big(1-\Delta Z^+(p)\big),
\end{equation}
with:
\begin{eqnarray} 
\label{5e8a}
\big(\Delta m_{\pi^+}^2\big)_{\pi\gamma} &=& (\frac{3e^2}
{\varepsilon}+\frac{7\alpha}{4\pi})m_{\pi}^2, \\
\label{5e8b}
-\Delta Z^+(p) &=& (2+\xi)\frac{e^2}{\varepsilon} 
+\frac{\alpha}{\pi} \bigg( 1+(1+\frac{\xi}{2})\ln(\frac{m_{\pi}^2}
{\lambda^2}) + \frac{1}{2}(1+\frac{\xi}{2})\frac{\lambda^2}{p^2}
\ln(\frac{m_{\pi}^2}{\lambda^2})\bigg),
\end{eqnarray}
$\varepsilon$ and $\lambda^2$ being defined in Eqs. (\ref{5e5a}) and
(\ref{5e5aa}), respectively.
\par
We next pass to the vertex function (Fig. 3a). Omitting from its definition 
the $\pi^+\pi^-\rightarrow \pi^0\pi^0$ amplitude $A(s|t,u)$, which at the 
tree level does not depend on $t$ and $u$ [Eq. (\ref{1e1})], it is:
\begin{eqnarray} \label{5e8}
\Lambda(p_1,p_2) &=& ie^2\int \frac{d^4k}{(2\pi)^4}\frac{\big((2p_1-k).
(2p_2+k)-\xi (2p_1-k).k(2p_2+k).k/k^2\big)}
{\big((p_1-k)^2-m_{\pi}^2+i\epsilon\big) \big((p_2+k)^2-
m_{\pi}^2+i\epsilon\big) \big(k^2+i\epsilon\big)}\nonumber \\
&\equiv& ie^2 \left[-R^{(1)}(p_1)+(4p_1.p_2+p_2^2-m_{\pi}^2)
F^{(1)}(p_1,p_2,0)
+2p_1^{\mu}F_{\mu}^{(1)}(p_1,p_2,0)\right]\nonumber \\
& &-ie^2\xi\bigg[-p_1^{\mu}\frac{\partial}{\partial p_1^{\mu}}R^{(1)}(p_1)
+(m_{\pi}^2-p_2^2)2p_1^{\mu}F_{\mu}^{(2)}(p_1,p_2,0)\nonumber \\
& & \ \ \ \ \ \ \ \ \ \ - 2p_2^{\mu}F_
{\mu}^{(1)}(p_1,p_2,0)-\overline R(p_1,p_2)\bigg],\ \ \ 
\end{eqnarray}
with the definitions and expressions of the functions $R$, $\overline R$ 
and $F$ given in Appendix D.
\par
To this order of approximation, the strong vertex part of $\Lambda$ is a
scalar vertex and therefore $\Lambda$ satisfies an identity typical of
mass operators, obtained by differentiation with respect to $m_{\pi}^2$
\cite{lb}. It reads:
\begin{equation} \label{5e10}
\Lambda(p_1,-p_1) = \frac{\partial \Sigma(p_1)}{\partial m_{\pi}^2} .
\end{equation}
The unrenormalized $\Lambda$ at leading orders is given in Eq. (\ref{be2}).
\par
In passing to the scattering amplitude, one must associate with the vertex 
function the wave function renormalization factors of the
external particles (the square-root of $(1-\Delta Z^+)$ of Eq. (\ref{5e7})).
This leads to the combination:
\begin{equation} \label{5e13}
\Lambda(p_1,p_2) -\frac{1}{2}\Delta Z^+(p_1) -\frac{1}{2}\Delta Z^+(p_2)
= (\frac{3e^2}{\varepsilon}-\frac{\alpha}{2\pi}) + \frac{\alpha}{2} 
\frac{4p_1.p_2}{s}\sqrt{\frac{s}{b^2}} \arctan\sqrt{\frac{b^2}{-b_0^2}} .
\end{equation}
[We have kept dominant terms only.] Notice that the gauge parameter and the 
logarithmic functions have been cancelled out.
\par
With the above quantity we must associate the contribution of the constraint
diagram arising from the interference term between the strong interaction
amplitude and one-photon exchange diagram (Fig. 3b):
\begin{eqnarray} \label{5e16}
\Lambda^C(p_1,p_2) &=& ie^2 \frac{i}{2\sqrt{s}} \int \frac{d^3k^T}
{(2\pi)^3} \frac{\big((2p_1-k^T).(2p_2+k^T)-\xi (2p_1-k^T).k^T(2p_2+k^T)/
k^{T2}\big)} {\big((p_1-k^T)^2-m_{\pi}^2
+i\epsilon\big)\big(k^{T2}+i\epsilon\big)}\nonumber \\
&=& ie^2\left[-\overline R^C(p_1)+4p_1.p_2F^{(1)C}(p_1,0)+4p^{T\mu}
F_{\mu}^{(1)C} (p_1,0)\right] \nonumber \\
& & -ie^2\xi\left[2\lambda^2 p^{T\mu}F_{\mu}^{(2)C}(p_1,0) + 
2p^{T\mu}F_{\mu}^{(1)C}(p_1,0) - \overline R^C(p_1)\right],
\end{eqnarray}
the definitions and expressions of the functions $\overline R^C$ and $F^C$
being given in Eqs. (\ref{4e23})-(\ref{4e23b}) and Appendix D.
One finds:
\begin{equation} \label{5e16a}
\Lambda^C(p_1,p_2) = -\frac{\alpha}{2} \frac{4p_1.p_2}{s}\sqrt{\frac{s}
{b^2}} \arctan \sqrt{\frac{b^2}{-b_0^2}} .
\end{equation}
\par
Adding this term to the quantity in Eq. (\ref{5e13}) yields
the net effect of the unrenormalized radiative corrections on the potential
$V_{00,+-}$ (relative to 1):
\begin{equation} \label{5e17}
\Lambda(p_1,p_2)-\frac{1}{2}\Delta Z^+(p_1)-\frac{1}{2}\Delta Z^+(p_2) +
\Lambda^C(p_1,p_2) = (\frac{3e^2}{\varepsilon}-\frac{\alpha}{2\pi}), 
\end{equation}
which is a gauge invariant quantity. On the other hand,
the constraint diagram has cancelled the term proportional
to the arctan function present in $\Lambda$, which otherwise would give
a contribution of order $O(\alpha^0)$.
\par
We next consider the contributions of the $O(e^2p^2)$ counterterms present 
in the chiral effective lagrangian. The complete expression of the 
corresponding $SU(3)\times SU(3)$ lagrangian in the Feynman gauge (in the 
standard scheme, to which we stick henceforth) is given in Ref. \cite{u},
where the coefficients of the various terms are designated by $K_i$
($i=1,\ldots ,17$). These terms are of two categories. Those which are
counterterms to the $O(e^2p^0)$ lagrangian (\ref{5e21}) and the infinite
parts of which are proportional to $C$, and those which are counterterms 
to the radiative corrections of the pion-photon interaction. Since the 
$(O(e^2p^0)$ term is included in the lagrangian contributing to the
amplitude $\mathcal {M}^{str.+q\gamma}$, so must be done with the 
corresponding counterterms. Therefore we have to select from the 
$O(e^2p^2)$ lagrangian those terms the infinite parts of which are not 
proportional to the constant $C$. 
\par
The $O(e^2p^2)$ lagrangian in the $SU(2)\times SU(2)$ case was 
recently presented in Ref. \cite{mms} (in the Feynman gauge). Since we are 
treating in the present work the $SU(2)$ case, we use for our subsequent 
calculations the latter lagrangian, indicating, when necessary, the 
correspondence between the $SU(2)$ and $SU(3)$ coefficients.
The relevant part of this lagrangian for our problem is (in standard
notations):
\begin{eqnarray} \label{5e50}
\mathcal{L}^{(e^2p^2)}&=&
F_0^2\bigg\{k_2<QUQU^{\dagger}><\partial_{\mu}U\partial^{\mu}U^{\dagger}>
\nonumber \\
& &\ \ \ \ +k_3\Big(<U^{\dagger}\partial_{\mu}UQ><U^{\dagger}\partial^{\mu}
UQ>+<\partial_{\mu}UU^{\dagger}Q><\partial^{\mu}UU^{\dagger}Q>\Big)
\nonumber \\
& &\ \ \ \ +k_4<U^{\dagger}\partial_{\mu}UQ><\partial^{\mu}UU^{\dagger}Q>
+k_7<QUQU^{\dagger}><\chi U^{\dagger}+U\chi^{\dagger}>\nonumber \\
& &\ \ \ \
+k_8<(U^{\dagger}\chi-\chi^{\dagger}U)(U^{\dagger}QUQ-QU^{\dagger}QU)>
\nonumber \\
& &\ \ \ \ +k_{10}<Q^2><\partial_{\mu}U\partial^{\mu}U^{\dagger}>
+k_{11}<Q^2><\chi U^{\dagger}+U\chi^{\dagger}>\bigg\},
\end{eqnarray}
where $Q$ is the quark charge matrix.
The coefficients $k$ have the following decompositions:
\begin{equation} \label{5e31}
k_i=\kappa_i \overline \lambda + k_i^r(\mu),
\end{equation}
where $\overline \lambda$ has a pole in $d=4$ dimensions,
\begin{equation} \label{5e32}
\overline \lambda = \frac{1}{16\pi^2}\left(\frac{1}{d-4}-\frac{1}{2}
\big(\ln(4\pi)+\Gamma'(1)+1\Big)\right),
\end{equation}
and the coefficients $\kappa$ have the values\footnote{We thank M. Knecht 
for pointing out to us the presence of the factor $-Z$ in $\kappa_8$.}:
\begin{equation} \label{5e33}
\begin{array}{llll}
\kappa_2=2Z, & \ \ \ \kappa_3=-\frac{3}{4}, & \ \ \ 
\kappa_4=-2Z, & \ \ \ \kappa_7=\frac{1}{4}+2Z,\\
\kappa_8=\frac{1}{8}-Z, & \ \ \ \kappa_{10}=-\frac{27}{20}-\frac{Z}{5}, 
& \ \ \ \kappa_{11}=-\frac{1}{4}-\frac{Z}{5},& \ \ \ Z=\frac{C}{F_0^4}. 
\end{array}
\end{equation}
The $\mu$-dependence of the finite part $k_i^r(\mu)$ of $k_i$ is
fixed by the prescription that its sum with the $\ln(\mu^{-2})$ term of the
corresponding loop diagrams be $\mu$-independent; the corresponding
multiplicative coefficient is $\kappa_i/(32\pi^2)$.
The lagrangian (\ref{5e50}), together with the coefficients (\ref{5e33}),
is associated with the $SU(2)\times SU(2)$ version of the chiral effective
lagrangian of Ref. \cite{gl1}, in which the multiplicative factors have
been arranged so that the coefficients $l_i$ keep the same 
values as those of the $O(4)$ version [Eqs. (\ref{6e1})-(\ref{6e3})]. 
\par
The lagrangian (\ref{5e50}) takes the 
following form in the representation (\ref{5ea}), keeping only pion fields 
that contribute to the scattering problem to the present approximation:
\begin{eqnarray} \label{5e30}
\mathcal{L}^{(e^2p^2)} &=& e^2\bigg(k_2+\frac{10}{9}(\frac{k_2}{10}+k_{10})
\bigg)\bigg[\partial_{\mu}\pi^0\partial^{\mu}\pi^0+2\partial_{\mu}\pi^+
\partial^{\mu}\pi^- \nonumber \\
& &\ \ \ \ \ \ \ \ \ \ + \frac{1}{F_0^2}\Big(\frac{1}{2}\partial_
{\mu}\pi^{02}
+\partial_{\mu}(\pi^+\pi^-)\Big) \Big(\frac{1}{2}\partial^{\mu}\pi^{02}
+\partial^{\mu}(\pi^+\pi^-)\Big)\bigg]\nonumber \\
& &-\frac{4e^2}{F_0^2}k_2\pi^+\pi^-\Big(\partial_{\mu}\pi^0\partial^
{\mu}\pi^0+2\partial_{\mu}\pi^+\partial^{\mu}\pi^-\Big)\nonumber \\
& &+\frac{e^2}{F_0^2}(2k_3-k_4)(\pi^+\stackrel{\leftrightarrow}
{\partial_{\mu}}\pi^-)(\pi^+\stackrel{\leftrightarrow}
{\partial^{\mu}}\pi^-) - \frac{e^2}{F_0^2}(2k_3+k_4)\bigg[F_0^2\partial_
{\mu}\pi^0\partial^{\mu}\pi^0\nonumber \\
& &\ \ \ \ \ -\pi^{02}\partial_{\mu}\pi^0\partial^{\mu}\pi^0
-2\pi^+\pi^-\partial_{\mu}\pi^0\partial^{\mu}\pi^0
+\frac{1}{2}\partial_{\mu}\pi^{02}\partial^{\mu}\pi^{02} + \partial_{\mu}
(\pi^{02})\partial^{\mu}(\pi^+\pi^-)\bigg]\nonumber \\
& &-\frac{5e^2}{9}(k_7+k_{11})8\hat mB_0\bigg[\frac{1}
{2}\pi^{02}+\pi^+\pi^- + \frac{1}{8F_0^2}(\pi^{02}+2\pi^+\pi^-)^2\bigg]
\nonumber \\
& &-\frac{2e^2}{F_0^2}\bigg(\frac{1}{4}(k_7-2k_8)+\frac{3}{4}(k_7+
2k_8)\bigg)8\hat mB_0\bigg[F_0^2\pi^+\pi^- 
- \pi^+\pi^-(\frac{1}{2}\pi^{02}+\pi^+\pi^-)\bigg].\nonumber \\
& &
\end{eqnarray}
The coefficients $(k_2/10+k_{10})$, $k_3$ and $(k_7+2k_8)$ contribute, 
at least concerning their infinite parts, only 
to the pion-photon interaction effects, while the coefficients 
$k_2$, $k_4$, $(k_7+k_{11})$ and $(k_7-2k_8)$ contribute to the
quark-photon interaction effects (insertions of the mass term (\ref{5e21})).
We shall admit that this separation also holds for their finite parts.
\par
The $O(e^2p^2)$ lagrangian (\ref{5e50}) introduces new terms into the
relationships between the pseudoscalar densities and the pion fields.
These are:
\begin{equation} \label{5e34}
P^a=2B_0F_0\Big(1+\frac{20e^2}{9}(k_7+k_{11})+4e^2k_8\delta_{a\pm}
-2e^2((k_7-2k_8)+3(k_7+2k_8))\frac{\pi^+\pi^-}{F_0^2}\Big)\pi^a,
\end{equation}
where $a=0,+,-$.
The three-pion term yields additional diagrams for the scattering
amplitude, which, however, contribute only off the mass shell. The term
proportional to the coefficient $k_8$ alone contributes to both the
pion-photon and quark-photon interaction effects. For this reason we
shall separate this coefficient into two parts by writing, with an 
obvious notation, $k_8=k_8^{\pi\gamma}+k_8^{q\gamma}$. Although this
separation is ambiguous for the corresponding finite parts, it will not 
show up in physical quantities.
\par
To obtain the renormalized Green's functions of the pseudoscalar densities
due to pion-photon interaction, we select among the above expreesions
the quantities proportional to the relevant coefficients. Furthermore,
since we are considering here on-mass shell expressions, the three-pion term 
of Eq. (\ref{5e34}) can be discarded.
One finds (in the Feynman gauge) the expressions: 
\begin{eqnarray} 
\label{5e35}
G^0(p) &=& \frac{i(2B_0F_0)^2 (1-\Delta Z^0)} {p^2-(m_{\pi}^2
+(\Delta m_{\pi^0}^2)_{\pi\gamma})+i\epsilon},\nonumber \\
-\Delta Z^0 &=& -\frac{2e^2}{9}(k_2^r+10k_{10}^r)+4e^2k_3^r,\nonumber \\
(\Delta m_{\pi^0}^2)_{\pi\gamma} &=& m_{\pi}^2 \bigg[-\frac{2e^2}{9}
(k_2^r+10k_{10}^r)+4e^2k_3^r\bigg],\\
& & \nonumber \\
\label{5e36}
G^+(p) &=& \frac{i(2B_0F_0)^2 (1-\Delta Z^+)} {p^2-(m_{\pi}^2
+(\Delta m_{\pi^+}^2)_{\pi\gamma})+i\epsilon},\nonumber \\
-\Delta Z^+ &=&  \frac{\alpha}{\pi}+\frac{\alpha}{\pi}\ln(\frac
{m_{\pi}^2}{\lambda^2})+\frac{\alpha}{2\pi}\frac{\lambda^2}{p^2}\ln(
\frac{m_{\pi}^2}{\lambda^2}-\frac{\alpha}{2\pi}\Big(\ln(\frac{m_{\pi}^2}
{\mu^2})+1\Big)\nonumber \\
& &\ \ \ -\frac{2e^2}{9}(k_2^r+10k_{10}^r)+8e^2k_8^{\pi\gamma,r},
\nonumber \\
(\Delta m_{\pi^+}^2)_{\pi\gamma} &=& m_{\pi}^2\bigg[\frac{7\alpha}{4\pi} 
- \frac{3\alpha}{4\pi}\Big(\ln(\frac{m_{\pi}^2}{\mu^2})+1\Big) - 
\frac{2e^2}{9}(k_2^r+10k_{10}^r) + 6e^2(k_7^r+2k_8^r)\bigg].\nonumber \\
& &
\end{eqnarray}
\par
The difference between the charged and neutral pion masses due to
pion-photon interaction is:
\begin{eqnarray} \label{5e37}
(\Delta m_{\pi}^2)_{\pi\gamma}&=&(\Delta m_{\pi^+}^2)_{\pi\gamma}-
(\Delta m_{\pi^0}^2)_{\pi\gamma}\nonumber \\
&=&\bigg[\frac{7\alpha}{4\pi}-\frac{3\alpha}{4\pi}\Big(\ln(\frac
{m_{\pi}^2}{\mu^2})+1\Big)-4e^2k_3^r+6e^2(k_7^r+2k_8^r)\bigg]m_{\pi}^2.
\end{eqnarray}
\par
To obtain the renormalized scattering amplitude of the process
$\pi^+\pi^-\rightarrow\pi^0\pi^0$, we consider the sum of 
the unrenormalized four-point vertex function and of the corresponding 
constraint diagram contribution:
\begin{equation} \label{5e53}
\bigg[W_{00,+-}+W_{00,+-}^{C(1,1,1)}\bigg]_{\pi\gamma}=(2B_0F_0)^{-4}
(1+4e^2k_8^{\pi\gamma})^{-2}A(s|t,u)(1+\Lambda+\Lambda^C),
\end{equation}
where $A(s|t,u)$ is defined in Eq. (\ref{1e1}), $\Lambda$ in Eq. 
(\ref{be2}) (taken here in the Feynman gauge) and $\Lambda^C$ in Eq.
(\ref{5e16a}); the other multiplicative factors come from the differences
between the pseudoscalar densities and the pion fields [Eqs. (\ref{5e34})].
The lagrangian (\ref{5e30}) provides the following counterterms:
\begin{eqnarray} \label{5e54}
(W_{00,+-}^{(e^2p^2)})_{\pi\gamma}&=&(2B_0F_0)^{-4}\frac{e^2}{F_0^2}
\bigg[\Big(\frac{2}{9}(k_2+10k_{10})-8k_3\Big)s\nonumber \\
& &\ \ \ \ \ \ \ \ \ \ +8\Big(k_3+\frac{3}{4}(k_7+2k_8)\Big)2\hat mB_0
\bigg].
\end{eqnarray}
These remove the divergences of the unrenormalized vertex (\ref{5e53}).
The sum of the two quantities is then multiplied by the renormalized
wave function renormalization factors $(2B_0F_0)^4(1-\Delta Z^0/2)^2
(1-\Delta Z^+/2)^2$ [Eqs. (\ref{5e35})-(\ref{5e36})]. We also incorporate 
in the mass term of the lowest-order amplitude (\ref{1e1}) the neutral
pion mass shift (\ref{5e35}).
The renormalized scattering amplitude, together with the constraint diagram 
contribution, is then: 
\begin{eqnarray} \label{5e39}
\bigg[\mathcal{M}_{00,+-}+\mathcal{M}_{00,+-}^{C(1,1,1)}\bigg]_{\pi\gamma} 
&=& \frac{(s-m_{\pi^0}^2)}{F_0^2}\bigg[1-\frac{\alpha}{2\pi}-
\frac{3\alpha}{4\pi}\Big(\ln(\frac{m_{\pi}^2}{\mu^2})+1\Big) - 4e^2k_3^r
\nonumber \\
& &\ \ \ -\frac{2e^2}{9}(k_2^r+10k_{10}^r)\bigg] + 2e^2\frac{m_{\pi}^2}
{F_0^2}\bigg[2e^2k_3^r+3e^2(k_7^r+2k_8^r)\bigg].\nonumber \\
\end{eqnarray}
The corresponding modification of the decay width is:
\begin{equation} \label{5e40}
(\Delta \Gamma)_{\pi\gamma}=\bigg[-\frac{\alpha}{\pi}-\frac{3\alpha}{2\pi}
\Big(\ln(\frac{m_{\pi}^2}{\mu^2})+1\Big)-\frac{4e^2}{9}(k_2^r+10k_{10}^r)
-\frac{16e^2}{3}k_3^r+4e^2(k_7^r+2k_8^r)\bigg]\Gamma_0.
\end{equation}
\par
To estimate the correction (\ref{5e40}) one needs to know the values of the
coefficients $k^r$. At this time they are not yet well established, but
several evaluations with different methods exist and can be used to have
an idea of the order of magnitude and sign of the correction. The
correspondence of the above coefficients $k$ with the $SU(3)$ coefficients
$K$ is (valid also for the finite parts):
\begin{eqnarray} \label{5e55}
& &k_2+10k_{10}=12K_1+10K_5,\ \ \ \ \ k_3=K_3,\nonumber \\
& &\frac{3}{4}(k_7+2k_8)=-\frac{3}{2}K_8+K_{10}+K_{11}.
\end{eqnarray}
\par
In ref \cite{bu}, the $K^r$'s have been evaluated using a resonance model
for the saturation of sum rules. The corresponding
values are in units of $10^{-3}$ and at the scale $\mu=m_{\rho}$: $K_1^r
=-6.4$, $K_3^r=6.4$, $K_5^r=19.9$, $K_8^r=K_9^r=K_{10}^r=0$, $K_{11}=0.6$.
\par
In Ref. \cite{bp} values of some of the $K^r$'s have been evaluated with the 
aid of the $1/N_C$-expansion method, the Extended Nambu--Jona--Lasinio model
and perturbative QCD. The results are not completely compatible with those
of Ref. \cite{bu}. In particular it is found (at similar scales and in 
units of $10^{-3}$): $K_9^r=-1.3$, $K_{10}^r=4.0$; the other results involve
$K^r$'s with even indices. Estimates of $K_{10}^r$ and $K_{11}^r$ have been
presented in Ref. \cite{m} on the basis of sum rule analysis. The value of
$K_{10}^r$ found there is compatible with that of Ref. \cite{bp}, while
for $K_{11}^r$ the value $2.9\times 10^{-3}$ is found (with the choices 
$\mu=\mu_0=m_V=m_A/\sqrt{2}$). The latter values, which affect the
combination $(k_7^r+2k_8^r)$, do not seem, however, to qualitatively change
the estimate of the correction (\ref{5e40}) evaluated with the values of 
the coefficients of Ref. \cite{bu}. We shall therefore present the numerical 
estimates with the values supplied by Ref. \cite{bu}.
\par
We find (with the mass scale $\mu=m_{\rho}$):
\begin{eqnarray} \label{5e41}
& & (\Delta m_{\pi^0})_{\pi\gamma}\simeq -0.01\ MeV,\ \ \ \ (\Delta m_{\pi^+}
)_{\pi\gamma}\simeq 0.43\ MeV,\ \ \ \ (\Delta m_{\pi})_{\pi\gamma}=0.44\
MeV,\nonumber \\
& & (\Delta \Gamma)_{\pi\gamma}=-0.20\alpha\Gamma_0=-0.0015\Gamma_0.
\end{eqnarray}
\par
We observe that the pion mass shift is of the order of 10\% of the
observed mass shift $\Delta m_{\pi}=4.6$ MeV and has the correct order of
magnitude for an $O(e^2p^2)$ effect.
The decay width correction is found almost equal to zero. The counterterm 
coefficients $k^r$ of Eq. (\ref{5e40}) give a negative contribution and 
thus have the tendancy to lower the value of $(\Delta \Gamma)_{\pi\gamma}$;
the contribution of the first two terms of the right-hand side of Eq.
(\ref{5e40}) is $0.0025\Gamma_0$.
\par

%\newpage

\section{Electromagnetic mass shift corrections}
\setcounter{equation}{0}

This section is devoted to the evaluation of the electromagnetic mass
shift corrections contained in the amplitude $\mathcal{R}e\mathcal{M}_{00,+-}
^{str.+q\gamma}$ [Eqs. (\ref{3e15})-(\ref{5e56})]. We shall proceed in two 
steps. First, we shall determine the effect of the insertions of the mass 
term (\ref{5e21}) in internal charged pion propagators. This will bring us 
to the strong interaction amplitude calculated with individual momenta fixed 
at the charged pion mass, but with the mass parameter fixed at the neutral 
pion mass. In the second step, we shall determine the difference between the 
latter amplitude and the one calculated with individual momenta and masses 
fixed at the charged pion mass, with which the numerical values of the 
scattering lengths are calculated in the literature \cite{bcegs,kmsf}.
\par
For the evaluation of the effect of the insertions of the mass shift term
(\ref{5e21}) the counterterm lagrangians needed here are the quark-photon 
part of the lagrangian (\ref{5e50}) and the standard $O(p^4)$ lagrangian 
\cite{gl1}; part of the latter survives in the mass shift counterterms. We
therefore begin by writing the relevant part of this lagrangian for our
problem (with standard notations):
\begin{eqnarray} \label{6e1}
\mathcal{L}^{(p^4)}&=&\frac{l_1}{4}<\partial_{\mu}U^{\dagger}\partial^{\mu}
U>^2+\frac{l_2}{4}<\partial_{\mu}U^{\dagger}\partial_{\nu}U><\partial^
{\mu}U^{\dagger}\partial^{\nu}U>\nonumber \\
& &+\frac{l_3}{16}<\chi^{\dagger}U+\chi U^{\dagger}>^2+\frac{l_4}{4}
<\partial_{\mu}\chi^{\dagger}\partial^{\mu}U+\partial_{\mu}\chi\partial^
{\mu}U^{\dagger}>.
\end{eqnarray}
The coefficients $l$ have the following decompositions:
\begin{equation} \label{6e2}
l_i=\gamma_i\overline{\lambda}+l_i^r(\mu),
\end{equation}
where $\overline{\lambda}$ is defined in Eq. (\ref{5e32}) and the 
coefficients $\gamma$ have the values:
\begin{equation} \label{6e3}
\gamma_1=\frac{1}{3},\ \ \ \ \gamma_2=\frac{2}{3},\ \ \ \ \gamma_3=
-\frac{1}{2},\ \ \ \ \gamma_4=2.
\end{equation}
For the scattering problem the above lagrangian becomes:
\begin{eqnarray} \label{6e4}
\mathcal{L}^{(p^4)}&=& \frac{l_1}{F_0^4}(\partial_{\mu}\pi^0\partial^{\mu}
\pi^0+2\partial_{\mu}\pi^+\partial^{\mu}\pi^-)(\partial_{\nu}\pi^0
\partial^{\nu}\pi^0+2\partial_{\nu}\pi^+\partial^{\nu}\pi^-)\nonumber \\
& &+\frac{l_2}{F_0^4}(\partial_{\mu}\pi^0\partial_{\nu}\pi^0+\partial_{\mu}
\pi^+\partial_{\nu}\pi^-+\partial_{\mu}\pi^-\partial_{\nu}\pi^+)
\nonumber \\
& & \ \ \ \ \ \ \ \ \ \ \times
(\partial^{\mu}\pi^0\partial^{\nu}\pi^0+\partial^{\mu}\pi^+\partial^{\nu}
\pi^-+\partial^{\mu}\pi^-\partial^{\nu}\pi^+)\nonumber \\
& &-\frac{l_3}{F_0^4}(2\hat mB_0)^2(\pi^{02}+2\pi^+\pi^-).
\end{eqnarray}
This lagrangian provides the following relationship between the 
pseudoscalar densities and the pion fields:
\begin{equation} \label{6e5}
P^a=(2B_0F_0)\Big(1+\frac{2l_3}{F_0^2}2\hat mB_0(1-\frac{\mbox{
\boldmath$\pi$}^2}{2F_0^2})-l_4\partial^2\Big)\pi^a.
\end{equation}
The three-pion term does not contribute to the present appreximation to 
the two-point Green's function renormalization, but provides an additional
contribution (a three-propa\-ga\-tor term) to the four-point Green's function.
Also, the term proportional to $l_3$ in Eq. (\ref{6e1}) provides in addition 
a two-propagator term to the four-point Green's function. These 
contributions should be taken into account for the off-mass shell
expression of the scattering amplitude.
\par
The finite parts of wave function and mass renormalizations proportional
to $C/F_0^4$ (in notations similar to those of Eqs. 
(\ref{5e35})-(\ref{5e36})) are:
\begin{eqnarray} 
\label{6e6}
-\Delta Z^0&=&-2e^2\bigg[k_2^r-k_4^r-\frac{20}{9}(k_7^r+k_{11}^r)\bigg],
\nonumber \\
(\Delta m_{\pi^0}^2)_{q\gamma}&=&-2e^2m_{\pi}^2\bigg[-\frac{C}{F_0^4}
\frac{1} {16\pi^2}\Big(\ln(\frac{m_{\pi}^2}{\mu^2})+1\Big)+k_2^r-k_4^r-
\frac{10}{9}(k_7^r+k_{11}^r)\bigg], \nonumber \\
& & \\
\label{6e7}
-\Delta Z^+&=&-2e^2\bigg[\frac{C}{F_0^4}\frac{1}{16\pi^2}\Big(\ln(\frac
{m_{\pi}^2}{\mu^2})+1\Big)+k_2^r-\frac{20}{9}(k_7^r+k_{11}^r) 
-4k_8^{q\gamma,r}-\frac{2C}{F_0^4}l_4^r\bigg],
\nonumber \\
(\Delta m_{\pi^+}^2)_{q\gamma}&=& 2e^2\frac{C}{F_0^2}-2e^2m_{\pi}^2\bigg[
\frac{2C}{F_0^4}\frac{1}{16\pi^2}\ln(\frac{m_{\pi}^2}{\mu^2})+k_2^r
-\frac{10}{9}(k_7^r+k_{11}^r)-(k_7-2k_8^r)\bigg].\nonumber \\
& & 
\end{eqnarray}
(We have also incorporated in $(\Delta m_{\pi^+}^2)_{q\gamma}$ the 
lowest order term (\ref{5e22})). The difference between the charged and
neutral pion masses due to quark-photon interaction is:
\begin{eqnarray} \label{6e8}
(\Delta m_{\pi}^2)_{q\gamma}&=&(\Delta m_{\pi^+}^2)_{q\gamma}-
(\Delta m_{\pi^0}^2)_{q\gamma}\nonumber \\
&=&2e^2\frac{C}{F_0^2}-2e^2m_{\pi}^2\bigg[\frac{C}{F_0^4}\frac{1}{16\pi^4}
\Big(3\ln(\frac{m_{\pi}^2}{\mu^2})+1\Big)+k_4^r-(k_7^r-2k_8^r)\bigg].
\end{eqnarray}
\par
To evaluate the change in the scattering amplitude due to the pion mass
shift (\ref{5e22}), we first calculate the pion loop contributions with
the physical pion masses in the internal propagators. There are diagrams
with tadpole loops at the four-pion vertex and four diagrams with loops 
with two-pion propagators, one with two charged pions, one with two neutral 
pions and two with one charged and one neutral pions. To the contributions
of these diagrams one adds the remaining contributions of the pseudoscalar
densities to the four-point vertex function [Eqs. (\ref{5e34}) and 
(\ref{6e5})].
The divergences of the above contributions are cancelled by those of the 
counterterm lagrangians (\ref{5e30}) (the part proportional to the
factor $C/F_0^4$) and (\ref{6e4}). One multiplies the result 
by the wave function renormalization factors $(1-\Delta Z^0/2)^2$
and $(1-\Delta Z^+/2)^2$ coming from the contributions 
(\ref{6e6})-(\ref{6e7}) and from those of the strong interaction limit
\cite{gl1} and isolates in the resulting
expression the part proportional to $C/F_0^4$. 
One also incorporates in the mass term of the lowest-order amplitude
(\ref{1e1}) the neutral pion mass shift (\ref{6e6}). One finds in the
limits $t=u=0$ (with the notations $m_{\pi}=m_{\pi^0}$ 
and $\Delta m_{\pi}^2=(\Delta m_{\pi}^2)_{q\gamma}$, Eq. (\ref{6e8})):
\begin{eqnarray} \label{6e9}
& &F_0^4\Big(\Delta \mathcal{R}e\mathcal{M}_{00,+-}\Big)_{q\gamma}=
\nonumber \\
& &\ \ \ \ \ \ -\frac{1}{32\pi^2}(s-m_{\pi}^2)
(s+p_1^2+p_2^2-2m_{\pi}^2)\Big(\frac{\Delta m_{\pi}^2}{m_{\pi}^2}
+\mathcal{R}e(Q^{+-}(s)-Q^{00}(s))\Big)\nonumber\\
& &\ \ \ \ \ \ +\frac{1}{16\pi^2}(s-m_{\pi}^2)(2-\mathcal{R}eQ^{00}(s))
\Delta m_{\pi}^2 -\frac{1}{64\pi^2}(p_1^2+p_2^2)(p_3^2+p_4^2)
\frac{\Delta m_{\pi}^2}{m_{\pi}^2}\nonumber \\
& &\ \ \ \ \ \ +2e^2F_0^2(3k_2^r-2k_4^r)s-2e^2F_0^2(2k_2^r-k_4^r)
(p_3^2+p_4^2) \nonumber \\
& &\ \ \ \ \ \ -2e^2F_0^2\Big(k_2^r-k_4^r-(k_7^r-2k_8^r)
\Big)m_{\pi}^2-2e^2F_0^2(2k_2^r-k_4^r)(s-m_{\pi}^2)\nonumber\\
& &\ \ \ \ \ \ +2e^2F_0^2(k_7^r-2k_8^r)(p_3^2+p_4^2-2m_{\pi}^2)
\nonumber\\
& &\ \ \ \ \ \ -4l_3^r(p_3^2+p_4^2-m_{\pi}^2)\Delta m_{\pi}^2-2l_4^r
(s-m_{\pi}^2)\Delta m_{\pi}^2.
\end{eqnarray}
(Details of the calculations can be found in Appendix E. $Q(s)$ is defined 
in Eq. (\ref{ee3}). Only relevant first-order terms in $\Delta m_{\pi}^2$ 
have been kept.)
\par
With the mass shell conditions $p_i^2=m_{\pi^+}^2$ ($i=1,\ldots,4$) and
$s=4m_{\pi^+}^2$, the above expression becomes, to first-order in 
$\Delta m_{\pi}^2$:
\begin{eqnarray} \label{6e10}
F_0^4\Big(\Delta \mathcal{R}e\mathcal{M}_{00,+-}(s=4m_{\pi^+}^2)\Big)_
{q\gamma}&=&\frac{11m_{\pi}^2}{16\pi^2}\Delta m_{\pi}^2
-(4l_3^r+6l_4^r)m_{\pi}^2\Delta m_{\pi}^2\nonumber \\
& &\ +2e^2F_0^2\Big((k_2^r-2k_4^r)+(k_7^r-2k_8^r)\Big)m_{\pi}^2.
\end{eqnarray}
\par
The relationships between the $k_i^r$'s and the $SU(3)$ coefficients 
$K_i^r$ \cite{u} are ($m_K$ is the kaon mass):
\begin{eqnarray}  \label{6e11}
k_2^r&=&K_2^r+K_6^r-\frac{1}{64\pi^2}\Big(\ln(\frac{m_K^2}{\mu^2})+1
\Big),\ \ \ \ \ \ k_4^r=-K_4^r,\nonumber \\
k_7^r+k_{11}^r&=&\frac{1}{5}(6K_8^r+5K_9^r+5K_{10}^r)-\frac{9}{320\pi^2}
\Big(\ln(\frac{m_K^2}{\mu^2})+1\Big),\nonumber\\
k_7^r-2k_8^r&=&10K_8^r-\frac{3}{16\pi^2}\Big(\ln(\frac{m_K^2}{\mu^2})
+1\Big).
\end{eqnarray}
\par
Numerical values for the coefficients $K_i^r$ are presented in Ref. 
\cite{bu}. They are in units of $10^{-3}$ and at the scale 
$\mu=m_{\rho}$: $K_2^r=-3.1$, $K_4^r=-6.2$, $K_6^r=8.6$, $K_8^r=K_9^r=
K_{10}^r=0$. The values of the coefficients $l_i^r$ 
can be found in Ref. \cite{bcegs}. They are in units of $10^{-3}$ and at 
the scale $\mu=m_{\rho}$: $l_1^r=-5.4$, $l_2^r=5.67$, $l_3^r=0.82$, 
$l_4^r=5.6$. One finds the following corrections in the scattering lengths
and the decay width:
\begin{equation} \label{6e14}
(\Delta (a_0^0-a_0^2))_{q\gamma}=0.0005,\ \ \ \ \ \
\frac{(\Delta \Gamma)_{q\gamma}}{\Gamma_0}=0.0035.
\end{equation}
\par
Once the correction due to the mass shift (\ref{5e22}) is isolated, 
the amplitude $\mathcal{R}e\mathcal{M}_{00,+-}^{str.+q\gamma}$ reduces
to the strong interaction amplitude calculated with individual momenta fixed 
at the charged pion mass [Eq. (\ref{5e56})] and the mass parameter fixed at
the neutral pion mass. Numerical values of the strong interaction
scattering lengths are calculated in the literature with the individual 
momenta and the mass parameter fixed at the charged pion mass 
\cite{bcegs,kmsf}. It is
therefore necessary to evaluate the difference between these two amplitudes.
It can be obtained from the $O(p^4)$ off-mass shell expression of 
the scattering amplitude given in Ref. \cite{gl1} (we neglect for the moment
the mass shift effect in the $O(p^6)$ term). Replacing in the $O(p^2)$
term (\ref{1e1}) the mass parameter in terms of $m_{\pi^0}^2$ with the 
use of the relation \cite{gl1}:
\begin{equation} \label{6e12a}
m_{\pi^0}^2=2\hat mB_0+\frac{m_{\pi}^4}{F_0^2}\Big(2l_3^r+\frac{1}
{32\pi^2}\ln(\frac{m_{\pi}^2}{\mu^2})\Big),
\end{equation}
one finds for the shift at threshold from the amplitude calculated with 
the charged pion mass:
\begin{eqnarray} \label{6e12}
\Big(\Delta \mathcal{R}e \mathcal{M}_{00,+-}(s=4m_{\pi^+}^2)\Big)_
{\Delta m_{\pi}}
&=&\frac{\Delta m_{\pi}^2}{F_0^2}\bigg[1+\frac{m_{\pi}^2}{F_0^2}\Big(
-\frac{9}{32\pi^2}-\frac{11}{16\pi^2}\ln(\frac{m_{\pi}^2}{\mu^2})
\nonumber \\
& &\ \ \ \ \ \ \ \ \ \ \ \ \ +4l_3^r+12l_4^r\Big)\bigg].
\end{eqnarray}
\par
Designating by $\mathcal{M}_{00,+-}^{str.}$ the strong interaction
amplitude (without mass shifts), one has at threshold [Eqs. (\ref{4e1}),
(\ref{4e2}) and (\ref{4e14})]:
\begin{equation} \label{6e13}
\mathcal{M}_{00,+-}^{str.}(s=4m_{\pi}^2)=\frac{32\pi}{3}(a_0^0-a_0^2)_
{str.},
\end{equation}
where the $a$'s are the scattering lengths calculated up to two loops
\cite{bcegs,kmsf} of the chiral effective lagrangian. In the standard 
scheme one has the value \cite{gl1,bcegs} : $(a_0^0-a_0^2)=0.258$, obtained
with charged pion masses. 
Eq. (\ref{6e12}) yields the following corrections for the scattering
lengths and the decay width
(we take $F_0=88$ MeV, $m_{\pi^+}=139.57$ MeV, $\Delta m_{\pi}=4.6$ MeV):
\begin{equation} 
\label{6e15}
(\Delta (a_0^0-a_0^2))_{\Delta m_{\pi}}=0.0083,\ \ \ \ \ \ 
\frac{(\Delta \Gamma)_{\Delta m_{\pi}}}{\Gamma_0}=0.064.
\end{equation}
The contribution of the $O(p^4)$ term represents 40\% of this correction, 
indicating an increase of its relative strength by a factor of 2 with
respect to the corresponding situation in the amplitude; this might be 
understood as a consequence of the increase of
the powers of mass and momemtum terms in higher-order terms. In this case
one should expect a correction of the order of $3\times 5\%$ coming from
the $O(p^6)$ term, which would bring the correction (\ref{6e15}) in the 
scattering lengths to 0.0095. 
\par
The effective scattering length corresponding to the amplitude
$\mathcal{R}e\mathcal{M}_{00,+-}^{str.+q\gamma}$ (\ref{3e15}) is then
(without the use of the $O(e^2p^4)$ correction coming from the $O(p^6)$
term):
\begin{equation} \label{6e15a}
(a_0^0-a_0^2)_{str.+q\gamma}=(a_0^0-a_0^2)_{str.(\pi^+)}+(\Delta
(a_0^0-a_0^2))_{q\gamma}+(\Delta (a_0^0-a_0^2))_{\Delta m_{\pi}}=0.267.
\end{equation}
(The inclusion of the estimated $O(e^2p^4)$ correction coming from the 
$O(p^6)$ term would bring this number to 0.268.)
\par
It is also possible to extract the value of the constant $C$ from
the pion mass shift formula (\ref{6e8}). Taking into account the mass 
shifts due to pion-photon interaction (0.4 MeV, Eqs. (\ref{5e37}) and 
(\ref{5e41})) and to isospin breaking (0.2 MeV \cite{gl1,gl2}) and the
observed pion mass difference (4.6 MeV), one finds $(\Delta m_{\pi})_
{q\gamma}\simeq$ 4.0 MeV, leading to:
\begin{equation} \label{6e16}
C=4.1\times 10^{-5} GeV^4.
\end{equation}
This value is in agreement with the central value found in Ref. \cite{bp}
at a slightly higher mass scale ($(4.2\pm 1.5)\times 10^{-5}$ GeV$^4$, at
$\mu =0.85$ GeV). The corresponding $O(e^2p^0)$ mass shift is $(\Delta
m_{\pi})_{q\gamma}^{(e^2p^0)}\simeq$ 3.6 MeV, the remaining 0.4 MeV
being produced by the $O(e^2p^2)$ correction [Eq. (\ref{6e8})], with
$(\Delta m_{\pi^0})_{q\gamma}=-0.13$ MeV.
\par

\section{Summary} 
\setcounter{equation}{0}

The total amount of  $O(\alpha)$ corrections to the lowest-order formula 
of the pionium decay width can be represented in the following form:
\begin{equation} \label{7e1}
\Gamma = \Gamma_0\sqrt{(1-\frac{\Delta m_{\pi}}{2m_{\pi^+}})}
\Big(1+\frac{(\Delta \Gamma)_{str.}}{\Gamma_0}\Big)
\Big(1+\frac{(\Delta \Gamma)_{\pi\gamma}}{\Gamma_0}\Big)
\Big(1+\frac{(\Delta \Gamma)_{q\gamma}}{\Gamma_0}\Big)
\Big(1+\frac{(\Delta \Gamma)_{\Delta m_{\pi}}}{\Gamma_0}\Big),
\end{equation}
where $\Gamma_0$ is the lowest-order decay width, Eq. 
(\ref{1e2}), with the 
strong interaction scattering lengths calculated up to two-loop order of 
the chiral effective lagrangian with charged pion masses; 
$(\Delta \Gamma)_{str.}$ is the correction
arising from the second-order perturbation theory contribution of the
interaction potential; $(\Delta \Gamma)_{\pi\gamma}$ 
arises from the radiative corrections due to pion-photon interaction;
$(\Delta \Gamma)_{q\gamma}$ is the correction coming from the pion mass
shift due to quark-photon interaction; $(\Delta \Gamma)_{\Delta m_{\pi}}$ 
is the correction coming from the shift in the strong interaction amplitude
due to the passage from the charged pion masses to the neutral pion 
masses with individual momenta fixed at the charged pion mass. The 
corresponding numerical values are (in the standard scheme):
\begin{eqnarray} \label{7e2}
\frac{(\Delta \Gamma)_{str.}}{\Gamma_0}&\simeq& 0.004, \ \ \ \ \ \ \
\frac{(\Delta \Gamma)_{\pi\gamma}}{\Gamma_0}\simeq -0.001,\nonumber \\
\frac{(\Delta \Gamma)_{q\gamma}}{\Gamma_0}&\simeq& 0.003, \ \ \ \ \ \ \ 
\frac{(\Delta \Gamma)_{\Delta m_{\pi}}}{\Gamma_0}\simeq 0.064.
\end{eqnarray}
These numbers are subject to uncertainties coming from the numerical 
values of the coefficints $k_i^r$ and $l_i^r$ of the counterterm 
lagrangians and from the higher loop effects not considered in the 
pion-photon and quark-photon interactions, corresponding to $O(e^2p^4)$
terms. These corrections should not qualitatively change the above
numerical estimates. One finds for the lifetime the value:
\begin{equation} \label{6e20}
\tau=2.97\times 10^{-15}\ \mathrm{s},
\end{equation}
to be compared with the value $\tau_0=3.19\times 10^{-15}$ s found from
formula (\ref{1e2}) using $(a_0^0-a_0^2)=0.258$.
\par
The evaluation of the corrections in the generalized scheme is more 
complicated because of the presence of additional terms in the counterterm 
lagrangians. However, their order of magnitude should not qualitatively
differ from those of the standard scheme.
\par

\vspace{1 cm}

\noindent
{\large \textbf{Acknowledgements}}
\par
\vspace{0.35 cm}

We are grateful to L. L. Nemenov for useful informations on the DIRAC
project. We thank M. Knecht, B. Moussallam and J. Stern for stimulating
discussions. One of us (H.S.) thanks J. Gasser and H. Leutwyler for
several helpful and clarifying discussions. He also enjoyed discussions
with E. A. Kuraev, P. Minkowski and A. G. Rusetsky.
\par

\newpage

\appendix
\renewcommand{\theequation}{\Alph{section}.\arabic{equation}}

\section{Integrals of the strong interaction constraint \protect \\
diagrams}
\setcounter{equation}{0}

This appendix is devoted to the calculation of the constraint
diagrams entering in the strong interaction corrections to the 
potentials (Sec. 4 and Fig. 1).
\par
The incoming particles are designated by 1 and 2, the outgoing ones by 3 
and 4, the loop particles by 5 and 6. They represent either $\pi^+\pi^-$
or $\pi^0\pi^0$. The relative momentum of the ingoing particles is
designated by $p$, that of the outgoing ones by $p'$. The loop particles 
have momenta $p_1-k$ and $p_2+k$. The constraint (\ref{2e1}) implies
that $p_{iL}=P_L/2$ ($i=1,\ldots ,6$) and $k_L=0$ inside the loop.
\par
The various momentum transfers have the following values:
\begin{eqnarray} \label{ae1}
t_{51}&=&(p_1-p_5)^2=k^{T2}, \ \ \ \ \ u_{61}=(p_1-p_6)^2=(2p^T-k^T)^2,
\nonumber \\
t_{35}&=&(p_5-p_3)^2=(p^{\prime T}-p^T+k^T)^2,\ \ \ \ \ \ u_{45}=
(p_5-p_4)^2=(p^{\prime T}+p^T-k^T)^2.\nonumber \\
& &
\end{eqnarray}
We also use notations (\ref{2e9}) with $b^2=-p^{T2}$ and $b^{\prime 2}
=-p^{\prime T2}$. We choose for the evaluation of the integrals values
of $s$ below $4m_{\pi}^2$, $m_{\pi}$ designating here the mass of the loop
pion, so that $b_0^2$ is negative. The analytic continuation of
$\sqrt{-b_0^2}$ to positive values of $b_0^2$ is done with the
replacement $\sqrt{-b_0^2}$ $\rightarrow$ $-i\sqrt{b_0^2}$. We use for
the integrations dimensional regularization.
\par
We introduce the compact notation:
\begin{equation} \label{ae2}
(a\widetilde g_0 b)=\int \frac{d^3k^T}{(2\pi)^3} a\widetilde g_0
(\frac{\sqrt{s}}{2},p^T-k^T)b,
\end{equation}
where $a$ and $b$ represent either one of the momenta $t$ and/or $u$ 
defined above or the factor 1. We find:
\begin{eqnarray}
\label{ae3}
(\widetilde g_0) &=& -\int \frac{d^3k^T}{(2\pi)^3} \frac{1}
{b_0^2+(p^T-k^T)^2+i\epsilon} = -\frac{1}{4\pi}\sqrt{-b_0^2}, \\
\label{ae4}
(\widetilde g_0 t_{51}) &=& (\widetilde g_0 u_{61}) =
\frac{1}{4\pi}(b^2+b_0^2)\sqrt{-b_9^2},\nonumber \\
(t_{35}\widetilde g_0) &=& (u_{45}\widetilde g_0) =
\frac{1}{4\pi}(b^{\prime 2}+b_0^2)\sqrt{-b_0^2}, \nonumber \\
(t_{35}\widetilde g_0 t_{51}) &=& (u_{45}\widetilde g_0 u_{61})
=-\frac{1}{4\pi}\bigg[ \frac{2}{3}tb_0^2+b^2b^{\prime 2}+\frac{5}{3}b_0^2
(b^2+b^{\prime 2})+(b_0^2)^2\bigg]\sqrt{-b_0^2},\nonumber \\
(t_{35}\widetilde g_0 u_{61}) &=& (u_{45}\widetilde g_0 t_{51})
=-\frac{1}{4\pi}\bigg[\frac{2}{3}ub_0^2+b^2b^{\prime 2}+\frac{5}{3}b_0^2
(b^2+b^{\prime 2})+(b_0^2)^2\bigg]\sqrt{-b_0^2}.\nonumber \\
& &
\end{eqnarray}
\par
Equations (\ref{ae4}) show that the integrals involving the variables
$t$ and/or $u$ yield, with $\widetilde g_{0,+-}$, quantities that are of 
order $O(\alpha^3)$ and higher and thus are negligible.
The similar terms considered with $\widetilde g_{0,00}$ yield 
contributions of order $O(\alpha^2\sqrt{\frac{\Delta m_{\pi}}{m_{\pi}}})$,
or $O((\frac{\Delta m_{\pi}}{m_{\pi}})^{3/2})$ and higher, which also
are imaginary. Apart from being small, these terms will cancel the
imaginary part of the amplitude present in the expression of the 
potential. The leading $O(\alpha)$ or $O(\sqrt{\frac{\Delta
m_{\pi}}{m_{\pi}}})$ terms of the constraint diagrams are simply obtained by
neglecting from the start the variables $t$ and $u$ in the scattering 
amplitudes, which then factorize outside the integrals and the latter 
reduce to the integral over the constraint propagator $\widetilde g_0$
[Eq.(\ref{ae3})].
\par

\section{Two-loop diagrams}
\setcounter{equation}{0}

In this appendix we generalize the validity of the lowest-order formula
established in Sec. 4 [Eq. (\ref{3e15})] to the two-loop level.
\par
The main ingredient of the proof is already met at the one-loop
level and can be summarized as follows. In the vicinity of the two-pion
threshold, the strong interaction unitarity one-loop amplitude (with the
pion mass shift included in) can be
decomposed into an analytic function $\mathcal {M}_{an.}^{(1)}$ of the 
real variables $b_0^2(s)$, $t$ and $u$ and a non-analytic part, $\mathcal{M}
_{nan.}^{(1)}$ (essentially proportional to $\sqrt{-b_0^2}$). [The 
analyticity of $\mathcal{M}_{an.}^{(1)}$ in $t$ and $u$ near threshold 
is due to the absence of infra-red singularities in the strong interaction 
amplitude with massive pions.] The constraint diagram amplitude, 
$\mathcal{M}^{C(1,0,1)}$ (the notation being explained before Eq. 
(\ref{4e22})), has the property of cancelling the nonanalytic part,
$\mathcal{M}_{nan.}^{(1)}$, of the one-loop amplitude (sum of the two
diagrams of Fig. 1):
\begin{equation} \label{bbe1}
\mathcal{M}^{(1)}+\mathcal{M}^{C(1,0,1)}=
\mathcal{M}_{an.}^{(1)}+\mathcal{M}_{nan.}^{(1)}+\mathcal{M}^{C(1,0,1)}
=\mathcal{M}_{an.}^{(1)}.
\end{equation}
The deviation of the analytic piece, $\mathcal{M}_{an.}^{(1)}$, from the
pionium energy to the $\pi^+\pi^-$ threshold is of order $O(\alpha^2)$
and hence $\mathcal{M}_{an.}^{(1)}$ can be replaced by its value at
threshold:
\begin{equation} \label{bbe2}
\mathcal{M}_{an.}^{(1)}(b_0^2,t,u)=\mathcal{M}_{an.}^{(1)}(s=4m_{\pi^+}^2,
t=0,u=0)+O(\alpha^2).
\end{equation}
Two cases must be distinguished, depending on whether the loop is made of
$\pi^+\pi^-$ or of $\pi^0\pi^0$. In the first case ($\pi^+\pi^-$-loop), 
the nonanalytic piece, $\mathcal{M}_{nan.}^{(1)}$, and the constraint 
amplitude, $\mathcal{M}^{C(1,0,1)}$, are real and separately vanish at the 
$\pi^+\pi^-$ threshold and therefore the value of $\mathcal{M}_{an.}^{(1)}$ 
at threshold coincides with that of the one-loop amplitude 
$\mathcal{M}^{(1)}$ at threshold, which is real. In the second case 
($\pi^0\pi^0$-loop), $\mathcal{M}_{nan.}^{(1)}$ and
$\mathcal{M}^{C(1,0,1)}$ are imaginary and do not vanish at the 
$\pi^+\pi^-$ threshold; the value of $\mathcal{M}_{an.}^{(1)}$ then 
coincides with the real part of the one-loop amplitude 
$\mathcal{M}^{(1)}$ at threshold. Therefore, in all cases we have:
\begin{equation} \label{bbe3}
\mathcal{M}^{(1)}+\mathcal{M}^{C(1,0,1)}=\mathcal{R}e\mathcal{M}^{(1)}
(s=4m_{\pi^+}^2,t=0,u=0).
\end{equation}
\par
This property can be used for the analysis of two-loop diagrams.
It is clear that not all two-loop diagrams necessitate a detailed study;
those not having singularities in the $s$-channel are analytic in the 
vicinity of the two-pion threshold and real; since in this case their 
deviation from the threshold value is of order $O(\alpha^2)$, they can 
immediately be replaced by their value at the $\pi^+\pi^-$ threshold.
\par
The typical diagrams involved in this analysis are presented in Fig. 4.
\par
The sum of the four diagrams of Fig. 4a (where the loop pions are the same 
in the four diagrams) is free of singularities in the $s$-channel and 
is represented in the vicinity of the two-pion threshold by a real analytic
function. The value of the latter at the pionium energy differs from its
value at the $\pi^+\pi^-$ threshold by an $O(\alpha^2)$ term; hence it can
be replaced by its value at the two-pion threshold. When the loops contain
at least one pair of $\pi^+\pi^-$, this value coincides with the real part
of the two-loop amplitude (first diagram) at threshold. If the two loops 
correspond to neutral pions, then this value differs from the real part of
the amplitude at threshold by a factor that is proportional to the product 
of the imaginary parts of each loop (actually cancelled by the last 
constraint diagram) and hence to $(\Delta m_{\pi})^2$. This factor
contributes with a relative order of magnitude of $10^{-4}$ and can be 
neglected. Therefore, the sum of the four diagrams at the pionium energy is 
equivalent to the real part of the two-loop amplitude (first diagram) at the 
$\pi^+\pi^-$ threshold.
\par
In diagrams of Fig. 4b, the tadpole diagram factorizes at the vertex and
does not interfere with the loop diagram calculations. For the latter, one
has the same results as in the one-loop case (Fig. 1).
\par
In the first diagram of Fig. 4c, the internal propagator is modified with
the inclusion of the self-energy correction (which must also be done 
in the other propagator). This feature does not
qualitatively change the singularities of the diagram with respect to the
one-loop case, for these depend essentially on the mass-shell condition
of the internal loops. A similar self-energy inclusion in the constraint
propagator [Eq. (\ref{2e7})] (second diagram) ensures the cancellation
of the singularities of the first diagram and the reasonings of the
one-loop case can be repeated. (The mass shift coming from the above
self-energy correction can be incorporated in the mass term used in the 
constraint propagator $\widetilde g_0$. Wave function renormalization 
constants should not influence physical results.)
\par
In the diagrams of Fig. 4d, the constraint diagram, corresponding to the 
first loop, cancels the singularities of the first loop of the first
diagram. The second loop, which is on the right of each diagram, is
free of singularities in the $s$-channel and is also free of infra-red
singularities (no massless particles). Hence, the sum of the two diagrams
is analytic in the vicinity of the two-pion threshold and the previous 
results are found again.
\par
There are also diagrams with three internal pion propagators, in which
three external pions join each other at one vertex and a single external
pion is attached at the other vertex. Such diagrams have singularities
for $p_i^2\ge(3m_{\pi})^2$ ($i=1,\ldots,4$). For $p_i^2$'s in the vicinity
of the mass shell, as is the case in the present problem, these 
diagrams are free of singularities and are analytic in $s,t,u$.
\par
The remaining two-loop diagrams do not have singularities in the 
$s$-channel and the above diagrams exhaust the cases where 
constraint diagrams occur. The result found in Sec. 4 
[Eq. (\ref{3e15})] can therefore be generalized to the two-loop level.
\par
At the three-loop level, a qualitative change appears with the occurrence
of inelasticities through the four-pion intermediate states. The constraint
propagator (\ref{2e7}) is no longer sufficient by itself to cancel the
singularities of the scattering amplitude. In this case, new pieces should
be added to it to take into account the inelasticity effects.
\par

\section{Cancellation of linear divergences}
\setcounter{equation}{0}

When the linear divergence of the continuous spectrum is kept, 
$(\Delta\Gamma)_{cont.}$ [Eq. (\ref{4e19})] receives the contribution:
\begin{equation} \label{ce1}
(\Delta\Gamma)_{cont.}=-2\Gamma_0\overline V_{+-,+-}\int_0^{\infty}
\frac{dk}{2\pi^2}+\cdots ,
\end{equation}
where the dots stand for the contributions other than that of the linear 
divergence.
\par 
On the other hand, the integration of the constraint propagator 
$\widetilde g_0$ [Eq. (\ref{2e7})] also produces a linear divergence.
Equation (\ref{ae3}) now becomes:
\begin{equation} \label{ce2}
(\widetilde g_0)=-\int\frac{d^3k^T}{(2\pi)^3}\frac{1}{b_0^2+
(p^T-k^T)^2+i\epsilon}=\int_0^{\infty}\frac{dk}{2\pi^2}-\frac{1}{4\pi}
\sqrt{-b_0^2}.
\end{equation}
The corresponding linear divergence shows up in the contributions of
constraint diagrams that enter in the calculation of the potentials.
We first consider the constraint diagram $\overline V_{+-,+-}\widetilde 
g_{0.+-} \widetilde T_{+-,00}$ (in lowest order) contributing to $V_{+-,00}$
in Eq. (\ref{4e6}). The linear divergence coming from $\widetilde
g_{0,+-}$ [Eq. (\ref{ce2})] gives an additional contribution to
$V_{+-,00}$:
\begin{equation} \label{ce3}
V_{+-,00}=\overline V_{+-,+-}V_{+-,00}\int_0^{\infty}\frac{dk}{2\pi^2}
+\cdots .
\end{equation}
This quantity, when inserted in the last term of the wave equation
(\ref{4e10}), provides, in first-order of perturbation theory, the 
contribution to the decay width:
\begin{equation} \label{ce4}
(\Delta\Gamma)^{C(2,0,2)}=2\Gamma_0 \overline V_{+-,+-}\int_0^{\infty}
\frac{dk}{2\pi^2},
\end{equation}
which cancels the linearly divergent contribution (\ref{ce1}).
In the last term of Eq. (\ref{4e10}) there are also quadratic terms in
$\int dk/(2\pi^2)$, which however are proportional to $\overline
V_{+-,+-}^2$. These must be associated with similar divergences arising
from third-order perturbation theory. The constraint diagrams of Fig. 2
considered in Sec. 5 do not contain linear divergences and hence the net
result of Eq. (\ref{4e28}) remains unchanged.
\par
The other linear divergences present in the constraint diagram 
contributions must similarly be associated with other contributions 
having the same types of divergences. Thus, the linear divergence of the
term $\overline V_{+-,+-}\widetilde g_{0,+-}\overline V_{+-,+-}$ which
enters in the calculation of $V_{+-,+-}$ [Eq. (\ref{4e6})] must be 
associated with the second-order perturbation theory contribution of
$\overline V_{+-,+-}$ of the wave equation (\ref{4e10}) which concerns 
the shift in the real part of the energy. The linear divergence of the
term $\frac{1}{2}\overline V_{+-,00}\widetilde g_{0,00}V_{00,+-}$,
contributing to $V_{+-,+-}$ cancels a similar divergence coming from the
last term of the wave equation (\ref{4e10}). The linear divergence of the
term $\frac{1}{2}V_{+-,00}\widetilde g_{0,00}V_{00,00}$ contributing to
the calculation of $V_{+-,00}$ must be associated with the contribution
of the last term of Eq. (\ref{4e8}) (and which was neglected in the text
because of the smallness of its finite part; cf. the paragraph preceding
Eq. (\ref{4e10})).
\par
In summary, the contributions of linear divergences, when isolated and
grouped in terms with definite structeres, disappear by mutual
cancellations from physical quantities. In this respect, the use of
dimensional regularization automatically takes into account from the start
the above cancellations.
\par

\section{Integrals of the radiative corrections}
\setcounter{equation}{0}

We present in this appendix the definitions and expressions of the integrals
involved in the electromagnetic radiative corrections.
For the evaluation of the divergent integrals we use dimensional
regularization and designate by $d$ and $\mu$ the space-time dimension and 
the regularization mass, respectively. 
\par
The integrals involved in the self-energy part are the following:
\begin{eqnarray} 
\label{5e3}
T(p) &=& \mu^{4-d}\int \frac{d^dk}{(2\pi)^d}\frac{1}{(p-k)^2-m_{\pi}^2+
i\epsilon} = im_{\pi}^2\big(\frac{1}{\varepsilon}+\frac{1}
{16\pi^2}\big),\\
\label{5e3a}
R_{\ ,\mu ,\mu\nu}^{(n)}(p) &=& \mu^{4-d}\int\frac{d^dk}{(2\pi)^d}\frac
{(1,k_{\mu},k_{\mu}k_{\nu})} {(p-k)^2-m_{\pi}^2+i\epsilon}
\frac{1}{(k^2+i\epsilon)^n},\\
\label{5e3b}
R^{(1)}(p) &=& \frac{i}{\varepsilon}+\frac{i}{16\pi^2}\bigg[2-
\frac{\lambda^2}{p^2}\ln(\frac{m_{\pi}^2}{\lambda^2})\bigg],\\
\label{5e4}
R_{\mu}^{(1)}(p) &=&  \frac{i}{2\overline \epsilon}p_{\mu} + 
\frac{i}{32\pi^2}p_{\mu}
\bigg[1-\frac{\lambda^2}{p^2}+(\frac{\lambda^2}{p^2})^2\ln(\frac
{m_{\pi}^2}{\lambda^2})\bigg],\\
\label{5e4a}
R_{\mu\nu}^{(2)}(p) &=& \frac{i}{4\overline \epsilon}g_{\mu\nu} +
\frac{i}{64\pi^2}g_{\mu\nu}\left(3+\frac{\lambda^2}{p^2}-\frac{2\lambda^2}
{p^2}\ln(\frac{m_{\pi}^2}{\lambda^2})-(\frac{\lambda^2}{p^2})^2\ln(\frac
{m_{\pi}^2}{\lambda^2})\right)\nonumber  \\
& & \ \ \ \ -\frac{i}{32\pi^2}\frac{p_{\mu}p_{\nu}}{p^2}\left(
\frac{\lambda^2+m_{\pi}^2}{p^2} - 2\frac{\lambda^2m_{\pi}^2}{(p^2)^2}
\ln(\frac{m_{\pi}^2}{\lambda^2})\right),
\end{eqnarray}
where we have defined:
\begin{eqnarray}
\label{5e5a}
\frac{1}{\varepsilon} &=& \frac{1}{16\pi^2}\bigg[\frac{1}{2-d/2}
-\ln(\frac{m_{\pi}^2}{\mu^2})+\Gamma'(1)+\ln(4\pi)\bigg],\\
\label{5e5aa}
\lambda^2 &=& m_{\pi^2}-p^2.
\end{eqnarray}
\par
One finds for the self-energy the expression:
\begin{eqnarray} \label{be1}
\Sigma(p) &=& \frac{e^2}{\varepsilon}(3m_{\pi}^2-(2+\xi)\lambda^2) 
+ \frac{\alpha}{\pi} \bigg[\frac{7}{4}m_{\pi}^2-\lambda^2 \nonumber \\
& & \ \ \ \ \ \ \ \ \ \ \ -(1+\frac{\xi}{2})\lambda^2 \ln(\frac
{m_{\pi}^2}{\lambda^2}) - \frac{1}{2}(1+\frac{\xi}{2})
\frac{(\lambda^2)^2}{p^2}\ln(\frac{m_{\pi}^2}{\lambda^2})\bigg].
\end{eqnarray}
\par
The integrals involved in the vertex function are the following:
\begin{eqnarray}
\label{5e9}
F_{\ ,\mu}^{(n)}(p_1,p_2,q) &=& \mu^{4-d}\int \frac{d^dk}{(2\pi)^d} 
\frac{(1,k_{\mu})} {\big((p_1-k)^2-m_{\pi}^2+i\epsilon\big)
\big((p_2+k)^2-m_{\pi}^2+i\epsilon\big) \big(k^2+i\epsilon\big)^n},
\nonumber \\
& & \\
\label{5e9b}
\overline R(p_1,p_2) &=& \mu^{4-d} \int \frac{d^dk}{(2\pi)^d}
\frac{1}{\big((p_1-k)^2-m_{\pi}^2+i\epsilon\big) \big((p_2+k)^2
-m_{\pi}^2+i\epsilon\big)}.
\end{eqnarray}
Their expressions are:
\begin{eqnarray} 
\label{5e11}
F^{(1)}(p_1,p_2,0) &=& -\frac{i}{8\pi s}\sqrt{\frac{s}{b^2}} \arctan
\sqrt{\frac{b^2}{-b_0^2}} \nonumber  \\
& & + \frac{i}{8\pi^2s}\bigg(2\ln(\frac{s}{\lambda^2}) - 4\ln2 + 4\bigg),
\ \ \ \ \ \ \ \ s\simeq 4m_{\pi}^2,\\
\label{5e11b}
F^{(1)}(p_1,-p_1,0) &=& -\frac{i}{16\pi^2}\frac{1}{p_1^2}\ln
(\frac{m_{\pi}^2}{\lambda^2}),\\
\label{5e11a}
F_{\mu}^{1)}(p_1,p_2,0) &=& -\frac{i}{16\pi^2s} p_{\mu}^T\left[ -4 + \pi
\sqrt{\frac{s}{b^2}}\bigg(\frac{\lambda^2}{b^2}\arctan \sqrt{\frac{b^2}
{-b_0^2}} - \sqrt{\frac{-b_0^2}{b^2}}\bigg)\right], \ \ \ \ s\simeq 
4m_{\pi}^2 ,\nonumber \\
& & \ \ \\
\label{5e11c}
F_{\mu}^{(1)}(p_1,-p_1,0) &=& -\frac{i}{32\pi^2}p_{1\mu}\left[ \frac{1}
{m_{\pi}^2} + \frac{1}{p_1^2} - 2\frac{\lambda^2}{(p_1^2)^2}\ln(\frac
{m_{\pi}^2}{\lambda^2}) + \frac{\lambda^2}{m_{\pi}^2p_1^2}\right],\\
\label{5e11d}
F_{\mu}^{(2)}(p_1,p_2,0) &=& \frac{i}{32\pi^2}(p_1-p_2)_{\mu}\frac{\pi}
{s^2}\left((\frac{s}{b^2})^{3/2}\arctan\sqrt{\frac{b^2}{-b_0^2}}-
\frac{s}{b^2} \sqrt{\frac{s}{-b_0^2}}\right),\ \ \ \ s\simeq 4m_{\pi}^2,
\nonumber \\
& & \\
\label{5e11e}
F_{\mu}^{(2)}(p_1,-p_1,0) &=& -\frac{i}{16\pi^2}\frac{p_{1\mu}}{p_1^2}
\left(-\frac{1}{\lambda^2}+\frac{1}{p_1^2}\ln(\frac{m_{\pi}^2}{\lambda^2})
\right), \\
\label{5e11f}
\overline R(p_1,p_2) &=& \frac{i}{\varepsilon} +\frac{2i}{16\pi^2}
\left(1-2\sqrt{\frac{-b_0^2}{s}}\big(\frac{\pi}{2}-\arctan\sqrt{\frac
{-4b_0^2}{s}}\big)\right), \\
\label{5e11g}
\overline R(p_1,-p_1) &=& \frac{i}{\varepsilon}.
\end{eqnarray}
One finds for the vertex function the expressions (at leading orders):
\begin{eqnarray} \label{be2}
\Lambda(p_1,p_2) &=& \frac{e^2}{\varepsilon}(3-(2+\xi)) + 
\frac{\alpha}{2\pi}
+ \frac{\alpha}{2} \frac{4p_1.p_2}{s} \sqrt{\frac{s}{b^2}} \arctan
\sqrt{\frac{b^2}{-b_0^2}}\nonumber \\
& & - \frac{\alpha}{2\pi} \frac{4p_1.p_2}{s}\bigg(2\ln(\frac{s}{\lambda^2})
-4\ln2 + 4\bigg) -\frac{\alpha \xi}{2\pi}\ln(\frac{m_{\pi}^2}{\lambda^2}),
\ \ \ \ \ s\simeq 4m_{\pi}^2,
\end{eqnarray}
\begin{equation} \label{be3}
\Lambda(p_1,-p_1) = \frac{e^2}{\varepsilon}(3-(2+\xi)) +
\frac{\alpha}{2\pi}(2+\xi)\bigg(1-\ln(\frac{m_{\pi}^2}{\lambda^2})-
\frac{\lambda^2}{p_1^2}\ln(\frac{m_{\pi}^2}{\lambda^2})\bigg).
\end{equation}
\par
Among the integrals involved in the calculation of the constraint
diagrams, $F^{(1)C}$ and $\overline R^C$ were given in Sec. 5 [Eqs.
(\ref{4e23})-(\ref{4e23b})]. The integral $F_{\mu}^{(2)C}$ is:
\begin{eqnarray} \label{be4}
F_{\mu}^{(2)C}(p_1) &=& \frac{i}{2\sqrt{s}}\int \frac{d^3k^T}{(2\pi)^3}
\frac{k_{\mu}^T}{\big((p_1-k^T)^2-m_{\pi}^2+i\epsilon\big)
\big(k^{T2}+i\epsilon\big)^2}\nonumber \\
&=& -\frac{1}{4\pi}\frac{i}{4\sqrt{s}}p_{\mu}^T\frac{1}{(b^2)^{3/2}}
\bigg( \arctan \sqrt{\frac {b^2}{-b_0^2}}-\frac{\sqrt{-b^2b_0^2}}{\lambda^2}
\bigg).
\end{eqnarray}
\par

\section{The scattering amplitude in the unequal mass case}
\setcounter{equation}{0}

In this appendix we present the details of the calculation of the 
off-mass shell $\pi^+(p_1)\pi^-(p_2)$ $\rightarrow$$\pi^0(p_3)\pi^0(p_4)$ 
scattering amplitude at $t=u=0$. We need for this evaluation the 
expression of the function
$\overline R$ [Eq. (\ref{5e9b})] in the unequal mass case. Defining:
\begin{equation} \label{ee1}
\overline R^{12}(s) = \mu^{4-d} \int \frac{d^dk}{(2\pi)^d}
\frac{1}{\big((p_1-k)^2-m_1^2+i\epsilon\big) \big((p_2+k)^2
-m_2^2+i\epsilon\big)},
\end{equation}
one has: 
\begin{eqnarray} \label{ee2}
\overline R^{12}(s)&=&\frac{i}{\varepsilon_1}+i\overline J^{12}(s),
\nonumber \\
\overline J^{12}(s)&=&\frac{1}{16\pi^2}
\bigg\{2-\frac{s-(m_1^2-m_2^2)}{2s}\ln(\frac{m_2^2}{m_1^2})-Q^{12}(s)
\bigg\},
\end{eqnarray}
with:
\begin{equation} \label{ee3}
Q^{12}(s)=\left\{
\begin{array}{l}
+\frac{\sqrt{4sb_0^2(s)}}{s}\Big[\ln\Big(\frac{\sqrt{s-(m_1-m_2)^2}
+\sqrt{s-(m_1+m_2)^2}}{\sqrt{s-(m_1-m_2)^2}-\sqrt{s-(m_1+m_2)^2}}\Big)
-i\pi\Big],\ \ \ \ (m_1+m_2)^2<s,\\
+\frac{\sqrt{-4sb_0^2(s)}}{s}\Big[\pi-\arctan\Big(\frac{\sqrt{-4sb_0^2(s)}}
{s-(m_1^2+m_2^2)}\Big)\Big],\ \ \ \ (m_1-m_2)^2<s<(m_1+m_2)^2,\\
-\frac{\sqrt{4sb_0^2(s)}}{s}\ln\Big(\frac{\sqrt{(m_1+m_2)^2-s}+\sqrt{
(m_1-m_2)^2-s}}{\sqrt{(m_1+m_2)^2-s}-\sqrt{(m_1-m_2)^2-s}},\ \ \ \
s<(m_1-m_2)^2.
\end{array}
\right.
\end{equation}
In Eq. (\ref{ee2}) $\varepsilon_1$ is equal to $\varepsilon$ [Eq. 
(\ref{5e5a})] with $m_1$ replacing $m_{\pi}$..
\par
For small negative values of $s$, denoted now by $t$, $\overline R^{12}$ 
behaves to first order in $\Delta m^2\equiv (m_1^2-m_2^2)$ as:
\begin{equation} \label{ee4}
\overline R^{12}(t)_{\stackrel{{\displaystyle =}}{t\rightarrow 0}}
\frac{i}{\varepsilon_1}+\frac{i}{16\pi^2}(\Delta m^2+\frac{t}{3})
\frac{2}{(m_1+m_2)^2}+O((\Delta m^2)^2,t\Delta m^2,t^2).
\end{equation}
\par
The loop contributions with $\pi^+\pi^-$ and $\pi^0\pi^0$ propagators,
respectively, are:
\begin{eqnarray} \label{ee5}
iF_0^4\mathcal{M}_{loop}^{+-}&=&\frac{1}{2}(s-2\hat mB_0)\bigg[(s+p_1^2+
p_2^2+2m_{\pi^+}^2-8\hat mB_0)\overline R^{+-}(s)+2T^+\bigg],\\
iF_0^4\mathcal{M}_{loop}^{00}&=&\frac{1}{2}(s-2\hat mB_0)\bigg[(p_3^2
+p_4^2+2m_{\pi^0}^2-6\hat mB_0)\overline R^{00}(s)+2T^0\bigg],
\end{eqnarray}
where $T^+$ and $T^0$ are defined in Eq. (\ref{5e3}), with the $\pi^+$
and $\pi^0$ masses, respectively.
\par
The sum of the two contributions with a $\pi^+\pi^0$ loop is, to first 
order in $\Delta m_{\pi}^2$ (regular terms in $t$ and $u$ are neglected):
\begin{eqnarray} \label{ee7}
iF_0^4\mathcal{M}_{loop}^{+0}&=&(p_1^2+p_2^2+2m_{\pi^+}^2-4\hat mB_0)T^+
+(p_3^2+p_4^2+2m_{\pi^0}^2-4\hat mB_0)T^0\nonumber \\
& &+\frac{1}{2}\Big[(p_1^2+m_{\pi^+}^2-2\hat mB_0)(p_4^2+
m_{\pi^0}^2-2\hat mB_0)\nonumber \\
& &\ \ \ \ \ \ +(p_2^2+m_{\pi^+}^2-2\hat mB_0)(p_3^2+m_{\pi^0}^2-2\hat mB_0)
\Big]\overline R^{+0}(t)\nonumber \\
& &+\frac{1}{2}\Big[(p_1^2+m_{\pi^+}^2-2\hat mB_0)(p_3^2+
m_{\pi^0}^2-2\hat mB_0)\nonumber \\
& &\ \ \ \ \ \ +(p_2^2+m_{\pi^+}^2-2\hat mB_0)(p_4^2+m_{\pi^0}^2-2\hat mB_0)
\Big]\overline R^{+0}(u)\nonumber \\
& &+\frac{1}{2}(p_1^2-p_3^2)(p_2^2-p_4^2)\overline R^{+0}(t)+
\frac{1}{2}(p_1^2-p_4^2)(p_2^2-p_3^2)\overline R^{+0}(u)\nonumber \\
& &+\frac{1}{2}(p_1^2+p_2^2-p_3^2-p_4^2)\Big[(\overline R^{+0}(t)+
\overline R^{+0}(u))\Delta m_{\pi}^2-(T^+-T^0)\Big]\nonumber \\
& &-\frac{1}{6}(p_1^2+p_2^2+p_3^2+p_4^2)\Big[(T^++T^0)
+(m_{\pi^+}^2+m_{\pi^0}^2)(\overline R^{+0}(t)+\overline R^{+0}(u)
\nonumber \\
& &\ \ \ \ \ +\frac{2i}{16\pi^2})\Big]
+\frac{1}{6t}(p_1^2-p_3^2)(p_2^2-p_4^2)\Big[(T^++T^0)\nonumber \\
& &\ \ \ \ \ -(m_{\pi^+}^2+m_{\pi^0}^2)(\overline R^{+0}(t)+\frac{i}
{16\pi^2})-t(\overline R^{+0}(t)-\frac{i}{48\pi^2})\Big]\nonumber \\
& &+\frac{1}{6u}(p_1^2-p_4^2)(p_2^2-p_3^2)\Big[(T^++T^0)\nonumber \\
& &\ \ \ \ \ -(m_{\pi^+}^2+m_{\pi^0}^2)(\overline R^{+0}(u)+\frac{i}
{16\pi^2})-u(\overline R^{+0}(u)-\frac{i}{48\pi^2})\Big].
\end{eqnarray}
\par
The tadpole diagrams at the four-pion vertex originate from the 
following part of the chiral effective lagrangian:
\begin{eqnarray}
\mathcal{L}_{tad.}&=&\frac{1}{2F_0^4}(\pi^{02}+2\pi^+\pi^-)(\frac{1}{2}
\partial_{\mu}\pi^{02}+\partial_{\mu}(\pi^+\pi^-))(\frac{1}{2}\partial
^{\mu}\pi^{02}+\partial^{\mu}(\pi^+\pi^-))\nonumber \\
& &-2\hat mB_0\frac{1}{16F_0^4}(\pi^{02}+2\pi^+\pi^-)^3
\end{eqnarray}
and yield the contribution:
\begin{eqnarray} \label{ee9}
iF_0^4\mathcal{M}_{tad.}&=&-\Big[3sT^0+4sT^+-(s-p_1^2-p_2^2-2m_{\pi^+}^2)
T^+\nonumber \\
& &\ \ \ \ -(s-p_3^2-p_4^2-2m_{\pi^0}^2)T^0\Big]+\frac{3}{2}2\hat mB_0
(3T^0+4T^+).
\end{eqnarray}
\par
The contribution to the four-point vertex function from the pseudoscalar
densities (\ref{5e34}) and (\ref{6e5}), concerning its mass shift
part, is:
\begin{eqnarray} \label{ee10}
F_0^4\Delta \mathcal{M}_P&=&-e^2F_0^2(\frac{80}{9}(k_7+k_{11})
+8k_8^{q\gamma})(s-2\hat mB_0)-4l_4\Delta m_{\pi}^2(s-2\hat mB_0)
\nonumber \\
& &+2e^2F_0^2(k_7-2k_8)(p_3^2+p_4^2-2m_{\pi^0}^2) 
-4l_3\Delta m_{\pi}^2(p_3^2+p_4^2-m_{\pi^0}^2).\nonumber \\
& &
\end{eqnarray}
\par
The $O(e^2p^2)$ lagrangian (\ref{5e30}) provides the contribution:
\begin{eqnarray} \label{ee11}
\mathcal{M}^{(e^2p^2)}&=&\frac{2e^2}{F_0^2}(3k_2-2k_4)s-\frac{2e^2}{F_0^2}
(2k_2-k_4)(p_3^2+p_4^2)\nonumber \\
& &-2e^2\frac{2\hat mB_0}{F_0^2}(\frac{10}{9}(k_7+k_{11})-(k_7-2k_8)).
\end{eqnarray}
\par

\newpage

\noindent
{\large \textbf{Figures}}

\newcounter{fig}
\begin{list}{Fig. \arabic{fig}.}
{\usecounter{fig}}
\item A loop diagram and its constraint diagram, denoted by a cross, in 
the $s$-channel. The sum of the two diagrams is free of singularities in 
the $s$-channel.
\item Constraint diagrams the dominant part of which cancel the ultra-violet 
divergence of the continuum contribution in second-order perturbation theory.
Wavy lines are photons.
\item One-photon exchange diagram in the presence of the strong coupling (a)
and its constraint diagram (b).
\item Two-loop diagrams having constraint diagram counterparts.
\end{list}
\par

\newpage

%%%%%%%%%%%%%%%%%%%%%%%%%%%%%%%%%%%%%%%%%%%%%%%%%%%%%%%%%%%%%%%%%%%%%%%%%%%%

%%%%%%%%%%%%%%% File of figures calling the FEYNMAN macropackage %%%%%%%%%%%

%%%%%%%%%%%%%%%%%%%%%%%%%%%%%%%%%%%%%%%%%%%%%%%%%%%%%%%%%%%%%%%%%%%%%%%%%%%%

%\documentclass[12pt]{article}
%\textheight 23 cm
%\textwidth 16 cm
%\oddsidemargin 0.2 cm
%\evensidemargin 0.2 cm
%\topmargin -1 cm
%\renewcommand{\baselinestretch}{1.4}

\input FEYNMAN

%\begin{document}

% page 1
\vspace*{1 cm}
\begin{picture}(40000,55000)
\THICKLINES

%%%%%%%%%%%%%%%%%%%%%%%%%%%%% Fig. 1 %%%%%%%%%%%%%%%%%%%%%%%%%%%%%%%%%%%% 

\put(9000,48000){\circle{4000}}
\drawline\fermion[\NW\REG](7000,48000)[4500]
\drawline\fermion[\SW\REG](\pfrontx,\pfronty)[4500]
\drawline\fermion[\NE\REG](11000,48000)[4500]
\drawline\fermion[\SE\REG](\pfrontx,\pfronty)[4500]

\put(18000,48000){\mbox{\boldmath $\leftarrow\ s\ \rightarrow$}}

\put(31000,48000){\circle{4000}}
\drawline\fermion[\NW\REG](29000,48000)[4500]
\drawline\fermion[\SW\REG](\pfrontx,\pfronty)[4500]
\drawline\fermion[\NE\REG](33000,48000)[4500]
\drawline\fermion[\SE\REG](\pfrontx,\pfronty)[4500]
\put(30500,47750){\mbox{\boldmath $\times$}}

\put(19000,39000){Fig. 1}

%%%%%%%%%%%%%%%%%%%%%%%%%%%%% Fig 2 %%%%%%%%%%%%%%%%%%%%%%%%%%%%%%%%%%%%%%%

\put(10000,25000){\circle{4000}}
\drawline\fermion[\NW\REG](8000,25000)[4500]
\global\advance\pbackx by -1200
\put(\pbackx,\pbacky){\mbox{\boldmath $\pi^0$}}
\drawline\fermion[\SW\REG](\pfrontx,\pfronty)[4500]
\global\advance\pbackx by -1200
\put(\pbackx,\pbacky){\mbox{\boldmath $\pi^0$}}
\drawline\fermion[\NE\REG](12000,25000)[4500]
\global\advance\pbackx by 200
\put(\pbackx,\pbacky){\mbox{\boldmath $\pi^+$}}
\drawline\fermion[\SE\REG](\pfrontx,\pfronty)[4500]
\global\advance\pbackx by 200
\put(\pbackx,\pbacky){\mbox{\boldmath $\pi^-$}}
\drawline\photon[\S\REG](10000,27000)[4]
\put(9900,27500){\mbox{\boldmath $\pi^+$}}
\put(9900,21500){\mbox{\boldmath $\pi^-$}}
\put(10500,24750){\mbox{\boldmath $\times$}}

\put(29000,25000){\circle{4000}}
\drawline\fermion[\NW\REG](27000,25000)[4500]
\global\advance\pbackx by -1200
\put(\pbackx,\pbacky){\mbox{\boldmath $\pi^0$}}
\drawline\fermion[\SW\REG](\pfrontx,\pfronty)[4500]
\global\advance\pbackx by -1200
\put(\pbackx,\pbacky){\mbox{\boldmath $\pi^0$}}
\drawline\fermion[\NE\REG](31000,25000)[4500]
\global\advance\pbackx by 200
\put(\pbackx,\pbacky){\mbox{\boldmath $\pi^+$}}
\drawline\fermion[\SE\REG](\pfrontx,\pfronty)[4500]
\global\advance\pbackx by 200
\put(\pbackx,\pbacky){\mbox{\boldmath $\pi^-$}}
\drawline\photon[\S\REG](29000,27000)[4]
\put(28900,27500){\mbox{\boldmath $\pi^+$}}
\put(28900,21500){\mbox{\boldmath $\pi^-$}}
\put(27700,24750){\mbox{\boldmath $\times$}}

\put(9500,18000){(a)}
\put(28500,18000){(b)}

\put(19500,12000){\circle{4000}}
\drawline\fermion[\NW\REG](17500,12000)[4500]
\global\advance\pbackx by -1200
\put(\pbackx,\pbacky){\mbox{\boldmath $\pi^0$}}
\drawline\fermion[\SW\REG](\pfrontx,\pfronty)[4500]
\global\advance\pbackx by -1200
\put(\pbackx,\pbacky){\mbox{\boldmath $\pi^0$}}
\drawline\fermion[\SW\REG](\pfrontx,\pfronty)[4500]
\global\advance\pbackx by -1200
\put(\pbackx,\pbacky){\mbox{\boldmath $\pi^0$}}
\drawline\fermion[\NE\REG](21500,12000)[4500]
\global\advance\pbackx by 200
\put(\pbackx,\pbacky){\mbox{\boldmath $\pi^+$}}
\drawline\fermion[\SE\REG](\pfrontx,\pfronty)[4500]
\global\advance\pbackx by 200
\put(\pbackx,\pbacky){\mbox{\boldmath $\pi^-$}}
\drawline\photon[\S\REG](19500,14000)[4]
\put(19400,14500){\mbox{\boldmath $\pi^+$}}
\put(19400,8500){\mbox{\boldmath $\pi^-$}}
\put(18200,11750){\mbox{\boldmath $\times$}}
\put(20000,11750){\mbox{\boldmath $\times$}}

\put(19000,5000){(c)}

\put(18500,500){Fig. 2}

\end{picture}

\newpage

% page 2
%\vspace*{1 cm}
\begin{picture}(40000,55000)
\THICKLINES

%%%%%%%%%%%%%%%%%%%%%%%%%%%%%% Fig. 3 %%%%%%%%%%%%%%%%%%%%%%%%%%%%%%%%%%%

\drawline\fermion[\NW\REG](9000,46000)[8000]
\global\advance\pbackx by - 1200
\put(\pbackx,\pbacky){\mbox{\boldmath $\pi^0$}}
\drawline\fermion[\SW\REG](9000,46000)[8000]
\global\advance\pbackx by - 1200
\put(\pbackx,\pbacky){\mbox{\boldmath $\pi^0$}}
\drawline\fermion[\NE\REG](9000,46000)[8000]
\global\advance\pbackx by 200
\put(\pbackx,\pbacky){\mbox{\boldmath $\pi^+$}}
\drawline\fermion[\SE\REG](9000,46000)[8000]
\global\advance\pbackx by 200
\put(\pbackx,\pbacky){\mbox{\boldmath $\pi^-$}}
\drawline\photon[\S\REG](13000,50000)[8]
\put(8500,37000){(a)}

\drawline\fermion[\NW\REG](30000,46000)[8000]
\global\advance\pbackx by - 1200
\put(\pbackx,\pbacky){\mbox{\boldmath $\pi^0$}}
\drawline\fermion[\SW\REG](30000,46000)[8000]
\global\advance\pbackx by - 1200
\put(\pbackx,\pbacky){\mbox{\boldmath $\pi^0$}}
\drawline\fermion[\NE\REG](30000,46000)[8000]
\global\advance\pbackx by 200
\put(\pbackx,\pbacky){\mbox{\boldmath $\pi^+$}}
\drawline\fermion[\SE\REG](30000,46000)[8000]
\global\advance\pbackx by 200
\put(\pbackx,\pbacky){\mbox{\boldmath $\pi^-$}}
\drawline\photon[\S\REG](34000,50000)[8]
\put(32000,45750){\mbox{\boldmath $\times$}}
\put(29500,37000){(b)}

\put(19000,33000){Fig. 3}

%%%%%%%%%%%%%%%%%%%%%%%%%%% Fig. 4a %%%%%%%%%%%%%%%%%%%%%%%%%%%%%%%%%%%%%%%%%%

\put(7000,20000){\circle{3500}}
\put(10500,20000){\circle{3500}}
\drawline\fermion[\NW\REG](5250,20000)[3500]
\drawline\fermion[\SW\REG](\pfrontx,\pfronty)[3500]
\drawline\fermion[\NE\REG](12250,20000)[3500]
\drawline\fermion[\SE\REG](\pfrontx,\pfronty)[3500]

\put(27000,20000){\circle{3500}}
\put(30500,20000){\circle{3500}}
\drawline\fermion[\NW\REG](25250,20000)[3500]
\drawline\fermion[\SW\REG](\pfrontx,\pfronty)[3500]
\drawline\fermion[\NE\REG](32250,20000)[3500]
\drawline\fermion[\SE\REG](\pfrontx,\pfronty)[3500]
\put(26500,19750){\mbox{\boldmath $\times$}}

\put(7000,9000){\circle{3500}}
\put(10500,9000){\circle{3500}}
\drawline\fermion[\NW\REG](5250,9000)[3500]
\drawline\fermion[\SW\REG](\pfrontx,\pfronty)[3500]
\drawline\fermion[\NE\REG](12250,9000)[3500]
\drawline\fermion[\SE\REG](\pfrontx,\pfronty)[3500]
\put(10000,8750){\mbox{\boldmath $\times$}}

\put(27000,9000){\circle{3500}}
\put(30500,9000){\circle{3500}}
\drawline\fermion[\NW\REG](25250,9000)[3500]
\drawline\fermion[\SW\REG](\pfrontx,\pfronty)[3500]
\drawline\fermion[\NE\REG](32250,9000)[3500]
\drawline\fermion[\SE\REG](\pfrontx,\pfronty)[3500]
\put(26500,8750){\mbox{\boldmath $\times$}}
\put(30000,8750){\mbox{\boldmath $\times$}}

\put(18500,1000){(a)}

\end{picture}

\newpage

% page 3
%\vspace*{1 cm}
\begin{picture}(40000,55000)
\THICKLINES

%%%%%%%%%%%%%%%%%%%%%%%%%%%%%%%%% Fig. 4b %%%%%%%%%%%%%%%%%%%%%%%%%%%%%%%%

\put(10000,48000){\circle{4000}}
\drawline\fermion[\NW\REG](8000,48000)[4500]
\drawline\fermion[\SW\REG](\pfrontx,\pfronty)[4500]
\drawline\fermion[\NE\REG](12000,48000)[4500]
\drawline\fermion[\SE\REG](\pfrontx,\pfronty)[4500]
\put(8750,48000){\circle{1500}}

\put(29000,48000){\circle{4000}}
\drawline\fermion[\NW\REG](27000,48000)[4500]
\drawline\fermion[\SW\REG](\pfrontx,\pfronty)[4500]
\drawline\fermion[\NE\REG](31000,48000)[4500]
\drawline\fermion[\SE\REG](\pfrontx,\pfronty)[4500]
\put(28500,47750){\mbox{\boldmath $\times$}}
\put(27750,48000){\circle{1500}}

\put(19000,40000){(b)}

%%%%%%%%%%%%%%%%%%%%%%%%%%%%%%%%% Fig. 4c %%%%%%%%%%%%%%%%%%%%%%%%%%%%%%%%

\put(10000,30000){\circle{4000}}
\drawline\fermion[\NW\REG](8000,30000)[4500]
\drawline\fermion[\SW\REG](\pfrontx,\pfronty)[4500]
\drawline\fermion[\NE\REG](12000,30000)[4500]
\drawline\fermion[\SE\REG](\pfrontx,\pfronty)[4500]
\put(10000,32750){\circle{1500}}

\put(29000,30000){\circle{4000}}
\drawline\fermion[\NW\REG](27000,30000)[4500]
\drawline\fermion[\SW\REG](\pfrontx,\pfronty)[4500]
\drawline\fermion[\NE\REG](31000,30000)[4500]
\drawline\fermion[\SE\REG](\pfrontx,\pfronty)[4500]
\put(28500,29750){\mbox{\boldmath $\times$}}
\put(29000,32750){\circle{1500}}

\put(19000,22000){(c)}

%%%%%%%%%%%%%%%%%%%%%%%%%%%%%% Fig. 4d %%%%%%%%%%%%%%%%%%%%%%%%%%%%%%%%%%5

\drawline\fermion[\W\REG](8000,10000)[4000]
\drawline\fermion[\SW\REG](\pfrontx,\pfronty)[4000]
\drawline\fermion[\NE\REG](\pfrontx,\pfronty)[8000]
\drawline\fermion[\E\REG](\pfrontx,\pfronty)[8000]
\put(12900,12100){\circle{4000}}

\drawline\fermion[\W\REG](27000,10000)[4000]
\drawline\fermion[\SW\REG](\pfrontx,\pfronty)[4000]
\drawline\fermion[\NE\REG](\pfrontx,\pfronty)[8000]
\drawline\fermion[\E\REG](\pfrontx,\pfronty)[8000]
\put(31900,12100){\circle{4000}}
\put(28500,10600){\mbox{\boldmath $\times$}}

\put(19000,4000){(d)}

\put(18500,100){Fig. 4}

\end{picture}

%%%%%%%%%%%%%%%%%%%%%%%%%% End of the file of figures %%%%%%%%%%%%%%%%%%%%%%


\newpage

\begin{thebibliography}{80}

\bibitem{w1}S. Weinberg, Phys. Rev. Lett. 17 (1966) 616.
\bibitem{w2}S. Weinberg, Physica A 96 (1979) 327.
\bibitem{gl1}J. Gasser and H. Leutwyler, Ann. Phys. (N.Y.) 158 (1984)
142.
\bibitem{gl2}J. Gasser and H. Leutwyler, Nucl. Phys. B250 (1985) 465.
\bibitem{p}J. L. Petersen, CERN Report No 77-04, 1975-1976
(unpublished); Phys. Reports C 2 (1971) 155.
\bibitem{fp}C. D. Froggatt and J. L. Petersen, Nucl. Phys. B129 (1977)
89.
\bibitem{sch}A. Schenk, Nucl. Phys. B363 (1991) 97.
\bibitem{em}P. Estabrooks and A. D. Martin, Nucl. Phys. B79 (1974) 301.
\bibitem{hh}B. Hyams \textit{et al.}, Nucl. Phys. B64 (1973) 134;
W. Hoogland \textit{et al., ibid.} B69 (1974) 266.
\bibitem{o}W. Ochs, thesis, Ludwig-Maximilians-Universit\"at, 1973.
\bibitem{r}L. Rosselet \textit{et al.}, Phys. Rev. D 15 (1977) 574.
\bibitem{nd}M. M. Nagels \textit{et al.}, Nucl. Phys. B147 (1979) 189;
O. Dumbrajs \textit{et al.}, Nucl. Phys. B216 (1983) 277.
\bibitem{g}J. Gasser, The Second DA$\Phi$NE Physics Handbook, eds.
L. Maiani, N. Paver and G. Pancheri (INFN-LNF publication, 1995), p. 215.
\bibitem{mp}D. Morgan and M. R. Pennington, The Second DA$\Phi$NE
Physics Handbook, eds. L. Maiani, N. Paver and G. Pancheri (INFN-LNF
publication, 1995), p. 193.
\bibitem{bcegs}J. Bijnens, G. Colangelo, G. Ecker, J. Gasser and M. Sainio,
Phys. Lett. B 374 (1996) 210.
\bibitem{dir}G. Czapek \textit{et al.}, Letter of intent, CERN/SPSLC 
92-44.
\bibitem{nm}L. L. Nemenov, Sov. J. Nucl. Phys. 41 (1985) 629.
\bibitem{af}L. G. Afanasyev \textit{et al.}, Phys. Lett. B 308 (1993)
200; B 338 (1994) 478.
\bibitem{fss}N. H. Fuchs, H. Sazdjian and J. Stern, Phys. Lett. B 269
(1991) 183.
\bibitem{gmorgw}M. Gell--Mann, R. J. Oakes and B. Renner, Phys. Rev.
175 (1968) 2198;
S. Glashow and S. Weinberg, Phys. Rev. Lett. 20 (1968) 224.
\bibitem{ssf}J. Stern, H. Sazdjian and N. H. Fuchs, Phys. Rev. D 47
(1993) 3814.
\bibitem{kms}M. Knecht, B. Moussallam and J. Stern, The Second DA$\Phi$NE
Physics Handbook, eds. L. Maiani, N. Paver and G. Pancheri (INFN-LNF
publication, 1995), p. 221.
\bibitem{kmsf}M. Knecht, B. Moussallam, J. Stern and N. H. Fuchs,
Nucl. Phys. B 457 (1995) 513; B 471 (1996) 445.
\bibitem{dgbth}S. Deser, M. L. Goldberger, K. Baumann and W. Thirring,
Phys. Rev. 96 (1954) 774.
\bibitem{up}J. L. Uretsky and T. R. Palfrey, Jr., Phys. Rev. 121 (1961)
1798.
\bibitem{t}T. L. Trueman, Nucl. Phys. 26 (1961) 57.
\bibitem{bnnt}S. M. Bilen'kii, Nguyen Van Hieu, L. L. Nemenov and
F. G. Tkebuchava, Sov. J. Nucl. Phys. 10 (1969) 469.
\bibitem{eil}G. V. Efimov, M. A. Ivanov and V. E. Lyobovitskii, Sov. J.
Nucl. Phys. 44 (1986) 296.
\bibitem{bpt}A. A. Bel'kov, V. N. Pervushin and F. G. Tkebuchava, Sov. J.
Nucl. Phys. 44 (1986) 300.
\bibitem{mrw}U. Moor, G. Rasche and W. S. Woolcock, Nucl. Phys. A 587
(1995) 747; A. Gashi, G. Rasche, G. C. Oades and W. S. Woolcock, preprint
IFA-97/13, nucl-th/9704017.
\bibitem{skg}M. Sander, C. Kuhrts and H. V. von Geramb, Phys. Rev. C
53 (1996) 2610.
\bibitem{lr}V. Lyubovitskij and A. Rusetsky, Phys. Lett. B 389 (1996)
181; V. E. Lyubovitskij, E. Z. Lipartia and A. G. Rusetsky, preprint
QFT-TSU/97-50, hep-ph/9706244.
\bibitem{k}E. A. Kuraev, preprint hep-ph/9702327.
\bibitem{ct}Ph. Droz-Vincent, Lett. Nuovo Cimento  1 (1969) 839; 
T. Takabayasi, Prog. Theor. Phys. 54 (1975) 563;
I. T. Todorov, Dubna Report No. E2-10125, 1976 (unpublished);
A. Komar, Phys. Rev. D 18 (1978) 1881;
H. Leutwyler and J. Stern, Phys. Lett. 73B (1978) 75;
H. W. Crater and P. Van Alstine, Ann. Phys. (N.Y.) 148 (1983) 57;
H. Sazdjian, Phys. Rev. D 33 (1986) 3401;
A. N. Mitra and I. Santhanam, Few-body Systems 12 (1992) 41;
G. Longhi and L. Lusanna, eds., Constraint's Theory and Relativistic 
Dynamics, proceedings of the Firenze Workshop, 1986 (World Scientific, 
Singapore, 1987).
\bibitem{qp}A. A. Logunov and A. N. Tavkhelidze, Nuovo Cimento
29 (1963) 380;
R. Blankenbecler and R. Sugar, Phys. Rev. 142 (1966) 1051;
F. Gross, Phys. Rev. 186 (1969) 1448;
M. H. Partovi and E. L. Lomon, Phys. Rev. D 2 (1970) 1999;
R. N. Faustov, Theor. Math. Phys. 3 (1970) 478;
C. Fronsdal and R. W. Huff, Phys. Rev. D 3 (1971) 933;
I. T. Todorov, Phys. Rev. D 3 (1971) 2351;
V. B. Mandelzweig and S. J. Wallace, Phys. Lett. B 197 (1987) 469.
\bibitem{lp}G. P. Lepage, Phys. Rev. A 16 (1977) 863.
\bibitem{js}H. Jallouli and H. Sazdjian, Ann. Phys. (N.Y.) 253 (1997) 376.
\bibitem{sbgmln}E. E. Salpeter and H. A. Bethe, Phys. Rev. 84 (1951) 
1232;
M. Gell--Mann and F. Low, Phys. Rev. 84 (1951) 350;
N. Nakanishi, Suppl. Prog. Theor. Phys. 43 (1969) 1. 
\bibitem{bcm}R. Barbieri, M. Ciafaloni and P. Menotti, Nuovo Cimento
55A (1968) 701.
\bibitem{l}S. Love, Ann. Phys. (N.Y.) 113 (1975) 153.
\bibitem{bybrbr}G. T. Bodwin and D. R. Yennie, Phys. Reports 43
(1978) 267;
R. Barbieri and E. Remiddi, Nucl. Phys. B141 (1978) 413;
W. Buchm\"uller and E. Remiddi, Nucl. Phys. B162 (1980) 250.
\bibitem{n}A. Nandy, Phys. Rev. D  5 (1972) 1531.
\bibitem{dgmly}T. Das, G. S. Guralnik, V. S. Mathur, F. E. Low and
J. E. Young, Phys. Rev. Lett. 18 (1967) 759.
\bibitem{egpdr}G. Ecker, J. Gasser, A. Pich and E. de Rafael, Nucl.
Phys. B321 (1989) 311.
\bibitem{u}R. Urech, Nucl. Phys. B433 (1995) 234.
\bibitem{llf}L. D. Landau and E. M. Lifshitz, Quantum Mechanics
(Pergamon Press, London, 1958), p. 127.
\bibitem{bf}M. A. B. B\'eg and R. C. Furlong, Phys. Rev. D 31 (1985)
1370.
\bibitem{j}R. Jackiw, M. A. B. B\'eg Memorial Volume, eds. A. Ali and 
P. Hoodbhoy (World Scientific, Singapore, 1991), p. 25. 
\bibitem{lb}J. H. Lowenstein, Commun. Math. Phys. 24 (1971) 1;
L. S. Brown, Ann. Phys. (N.Y.) 126 (1980) 135. 
\bibitem{mms}Ulf-G. Meissner, G. M\"uller and S. Steininger, preprint
KFA-IKP(TH)-1997-06, hep-ph/9704377 (to appear in Phys. Lett. B).
\bibitem{bu}R. Baur and R. Urech, Phys. Rev. D 53 (1996) 6552;
preprint ZU-TH 30/96, hep-ph/9612328.
\bibitem{bp}J. Bijnens and J. Prades, preprint FTUV/96-69, 
hep-ph/9610360.
\bibitem{m}B. Moussallam, preprint IPNO/TH 97-03, hep-ph/9701400.

\end{thebibliography}
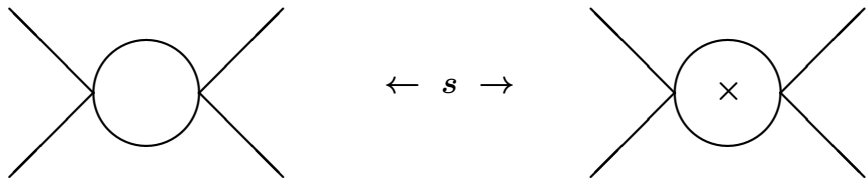
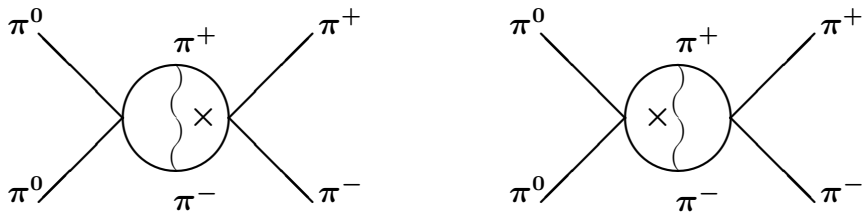
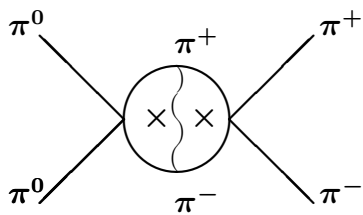
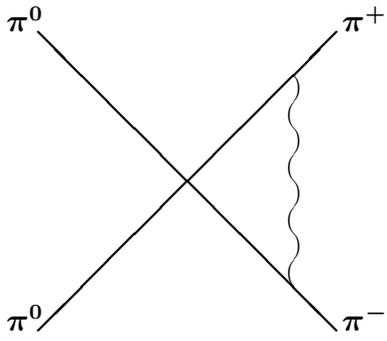
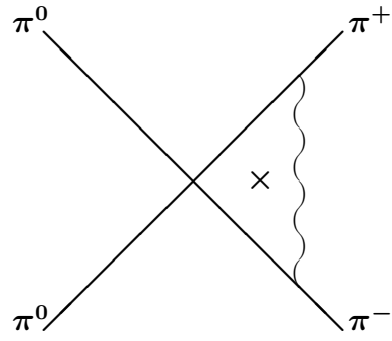
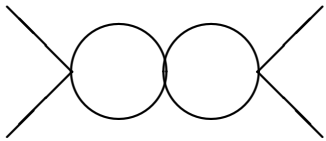
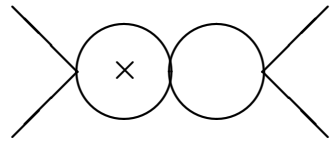
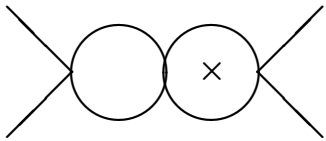
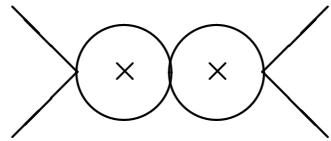
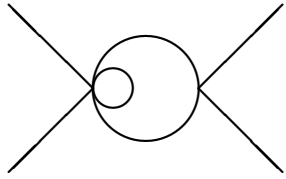
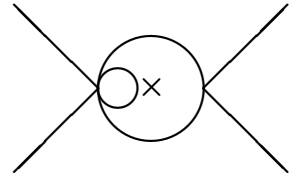
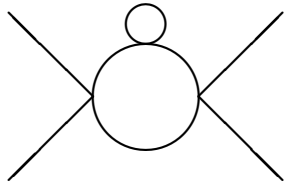
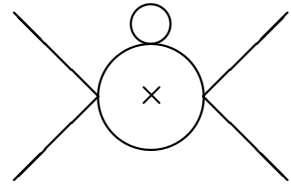
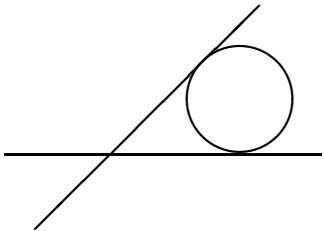
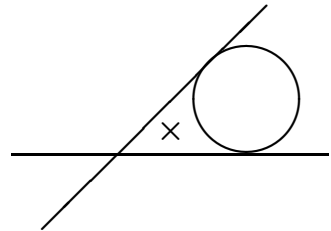
\end{document}